\documentclass[12pt]{article}

 \usepackage{amsfonts}
 \usepackage{amssymb,amsmath}
 \usepackage{graphicx} \usepackage{epstopdf}
  \numberwithin{equation}{section}
 \newcommand{\crlb}[1]{\label{#1}\\[2pt]}
 
 \newcommand{\crld}[1]{\label{#1}}
 \newcommand{\eela}[1]{\quad\hbox{\scriptsize{#1}}\label{#1}\end{eqnarray}}
 \newcommand{\eelb}[1]{\label{#1}\end{eqnarray}}
 
 \newcommand{\newsecb}[2]{\section{#1}\label{#2}\setcounter{equation}{0}}

 \newcommand{\nolabels} {\def\eel{\eelb}\def\eeql{\eeqlb}  \def\crl{\crlb} \def\newsecl{\newsecb}\def\bibiteml{\bibitem} \def\citel{\cite}\def\labell{\crld}}
\newcommand{\eeqla}[1]{\quad\hbox{\scriptsize{#1}}\label{#1}\end{aligned}\end{equation}}
\newcommand{\eeqlb}[1]{\label{#1}\end{aligned}\end{equation}}

\newcommand\publishversion{\nolabels\setlength{\textheight}{8.3in}\setlength{\oddsidemargin}{-.5in}
   	 \setlength{\textwidth}{7in}\setlength{\topmargin}{-0.2in}}
\def\beq{\begin{equation}\begin{aligned}}		\def\eeq{\end{aligned}\end{equation}}
\def\be{\begin{eqnarray}}  					\def\ee{\end{eqnarray}}		%\be and \ee may become obsolete in due time.
\def\bi#1{\begin{itemize}\item[#1]} 			 			\def\ei{\end{itemize}}
  \def\eqn#1{(\ref{#1})}
\def\Tr{{\mbox{Tr}}\,}   	 \def\fn{\footnote}	  		\def\nm{\nonumber}

		 \def\alf{\alpha}   \def\bet{\beta}   \def\del{\delta}        \def\lam{\lambda}
		       \def\rr{\varrho}      
    		  	\def\Del{\Delta}    \def\eps{\epsilon}
	    		        		     		\def\vv{\varphi}     
 	 		     	\def\tht{\theta}      	
	     	         \def\w{\omega}
\def\W{\Omega}    		  		\def\dd{{\rm d}} 		
 		     \def\ZZ{\mathbb{Z}}
\def\NN{\mathcal{N}} 		     
\def\pa{\partial}			\def\ra{\rightarrow}	
\def\bra{\langle} 		\def\ket{\rangle}

\def\qu{\overset{\textstyle ?}{=}}

\def\fract#1#2{{\textstyle\frac{#1}{#2}}}	 	 
\def\ffract#1#2{\raise .2 em\hbox{$\scriptstyle#1$}\kern-.3em/\kern-.2em\lower .15 em \hbox{$\scriptstyle#2$}}
\def\half{\fract12}					
\def\tl#1{\tilde{#1}}
\def\ex#1{e^{\textstyle#1}} 		\def\qqquad{\qquad\qquad}		\def\qqqquad{\qqquad\qqquad}
\def\bpmatrix{\begin{pmatrix}} 			\def\epmatrix{\end{pmatrix}}
\def\bmatrix{\begin{matrix}} 			\def\ematrix{\end{matrix}}
\def\bcenter{\begin{center}}			\def\ecenter{\end{center}}
\def\ret{\\[5pt]}  \def\ds{\displaystyle}
\def\inn{{\mathrm{in}}}  \def\outt{{\mathrm{out}}} 
\def\tott{{\mathrm{tot}}}
      \def\BH{{\mathrm{BH}}}     
\def\widthfig#1#2{\(\hbox{\includegraphics[width=#1]{#2}}\)}

\def\lowerheightfig#1#2#3{\(\raise-#1\hbox{\includegraphics[height=#2]{#3}}\)}
\def\lowerwidthfig#1#2#3{\(\raise-#1\hbox{\includegraphics[width=#2]{#3}}\)}
\def\th{\({}^{\mathrm{th}}\)}		\def\st{\({}^{\mathrm{st}}\)}		 
 \def\weglaten#1{ } \def\Ret{\\[13pt]} 

 				  \publishversion

\begin{document} 
\begin{center}

{\Large \textbf{Alternative theory for the quantum black hole\ret and the temperature of its quantum radiation}\fn{To appear as Chapter 3 in:
\emph{ The Black Hole Information Paradox}, A. Akil and C. Bambi, \\ editors (Springer Singapore).}}
\Ret
\large Gerard 't~Hooft\Ret \normalsize
Institute for Theoretical Physics \\ Utrecht University  \\[10pt]
Princetonplein 5 \\
3584 CC Utrecht \\
 the Netherlands\\[15pt] \end{center}
% \begin{titlepage} \maketitle\end{titlepage}
% For the book, it may be necessary to distinguish sections and subsections from chapters. In this chapter, I refer
% to the next layers as sections and subsections; this may be changed or adapted to the use in other chapters.
\textbf{\large Abstract} \begin{quote}
A mechanism is found that explains how matter falling into the future event horizon of a black hole leaves information there, which it sends to the past event horizon, and there it determines how particles are emitted. This way information must be conserved. The mechanism is a calculable gravitational effect. We also show how it is avoided that the `hidden region' of the black hole gets involved, which has been standing out as a problem ever since Hawking found that particles must be emitted. The most striking consequence of our mechanism is that the radiation temperature is not what Hawking calculated, but twice that value. This is a direct consequence of the fact that the `hidden region' is not there.\Ret\end{quote}
\textbf{\large Contents of this Chapter}
 \begin{quote}

 The Schwarzschild solution of Einstein's equations\,\cite{Schw.ref} 
is a smooth, non quantum mechanical exact solution, describing the periphery of a black hole, where Standard Model physics should still be applicable. We may be tempted to start with describing the situation with `no particles present', and call that the vacuum state. But  this definition turns out  not to be sufficiently precise. When an object such as an elementary particle falls into a black hole, it gradually becomes invisible, and similarly, the particles emitted by  a black hole, will  be difficult to follow far backwards in time. Yet we show that with only a few natural assumptions, one can set up equations that can be solved. We find that black holes produce an almost perfect thermal distribution of out-going particles.

However, there is an important point where many colleagues do not agree with the  author of this chapter. What is the \emph{temperature} of this radiation? We explain in this chapter how one can argue  that the temperature will be twice as high as mostly thought, by Stephen Hawking and his followers. This factor 2 emerges, when we continue the Schwarzschild solution analytically, as far as we can. There, one discovers an entirely new universe, as if there were two universes. One is ours, full of particles, while the other is not well understood. We insist to analyse this feature as accurately as we can, using equations that must make sense.

The temperature of the radiation emitted by a black hole can be calculated in several ways. The best way is to first calculate the entropy of a black hole, by meticulously counting the number of independent quantum states. This has to be done with care.  At every moment in time we only count the particle states that are visible in principle, so that these are the building blocks of all black hole states. Particles are visible if an observer far away can use his operators to create or annihilate that particle. We must count the orthonormal states,  as it is done in ordinary physics, not the entangled states, since parts of these are often invisible. Going to different values of the time parameter, this set of visible particles is continuously transformed into different sets, but we have no problems with that.
All this will be further explained.

By identifying the operators that create or annihilate  elementary particles near a black hole, one can follow the evolution of the system as a function of time. The operators far from the horizon are related to the operators close to the horizon, by Bogoljubov transformations. This implies that  an observer near the crossing point of future and past event horizon, sees particles differently from what the far-away observer sees. We calculate how these particles are related using the Bogoljubov transformation.

And then we apply a time translation in the future direction, for the distant observer. This transformation 
reduces the visibility of the in-going particles, but the out-particles become better visible. This transformation is called a `firewall transformation'. The most central process in our formalism uses the fact that removing an in-particle causes a shift in the wave functions of all out-particles. This shift has been calculated, it is basically the Shapiro effect for radiation grazing past a heavy object, due to the gravitational field. At early times, this shift may be of the order of the Planck scale, but it increases exponentially in time. The mathematics is basically very easy. The position of the emerging out-particle is identified with the momentum of the disappearing in-particle. And so it happens that, while an in-particle is annihilated, an out-particle is formed, in such a way that the out-wave function is simply a Fourier transform of the in-wave function.

This is how, as time passes, the set of visible particles continuously changes, exactly as one should have expected juist by staring at the physics.

But then there is one thing that still has to be worked out: \emph{Which of the two universes carries the in-particles, and in which of the two universes does the out-going particle go?} Remember that this
universe actually contains two universes; it is too big. Again there is a natural answer: \emph{the `hidden' universe is an exact copy of the visible universe.} \\ We explain why this is so.

A consequence of this is that the wave functions of all in-particles must br \emph{even} functions of their momenta \(p\),
	\be\psi^\inn(p)=\psi^\inn(-p)\ . \nm\eel{ppeven.eq}
The Fourier transform of this gives an out-wave function that is even as well:
	\be\psi^\outt(u)=\psi^\outt(-u)\ . \nm\eel{uueven.eq}
This implies that knowing \(\psi\) in one universe is all we need to obtain the Fourier transformed wave function in the other universe. They both remain even. The mathematics of `even Fourier transforms', or Fourier transforms on half-lines, is easy. Instead of the usual expression our momentum space in-wave functions \(\psi(p)\) for positive momenta \(p\), is transformed by the Fourier operator \(F^+\) as follows:
	\be F^+\psi(u)=\sqrt{\fract 2 {\pi}}\int_0^\infty\dd p\,\cos(p\,u)\,\psi(p)\ . \nm\ee %l{Fplus.eq}
This function is its own inverse. This, we claim, is our solution to the information loss problem.

	To see what the temperature of the black hole radiation is, we consider the time dependence of the visible particle states. One finds that the Bogoljubov transformation gives the time dependence of the states in the two universes, which we call region \(I\) and region \(II\), in terms of the Hamiltonian, 
	\be H^\tott=H_I-H_{II}\ . \nm\ee
This operator is exactly conserved, because, locally, it is the generator of the Lorentz transformations in the \(r,\,t\) direction, and because of exact Lorentz invariance its value starts out as zero, so that \(H^\tott=0\) always. The reason for that is that the black hole is switched on in the Schwarzschild metric, where the two universes are identical
 (This is not so in Rindler space, where one starts out with non-Lorentz invariant states, and this is why, in Rindler space, the treatment must  be different).
We see that the wave function used by distant observers will be
	\be| \W\ket=\sum_{\psi}|\psi\ket\,\bra\psi|\ , \nm\ee
where we use Dirac's bra-ket notation just to indicate that the states \(\bra\psi|\) evolve with minus the Hamiltonian used for the states \(|\psi\ket\).

This is an exact density matrix \(\rr\)  that evolves with the evolution law
	\be \dot\rr=i[\rr,H]\ , \nm\ee
From the Bogoljubov transformation, we find the density matrix \(\rr\) from the state \(|\W\ket\) as follows :
\be \bra n_{II}|\,\rr\, |\,n_{I}\ket=
\bra n_I,\,n_{II}|\W\ket=C^{st} \ex{-\pi n_I\,\w}\,  \del_{n_I,\,n_{II}} \nm\ee
where \(n\) is the number of particles with energy \(\w\), and \(C^{st}\) is a normalisation factor.
This gives a density matrix \(\ex{-\pi H}\) if \(H=n\,\w\)\,  and  \(\bet=1/T=\pi\):
	\be \rr= \ex{-\pi H}\ .\nm\ee
\(H\) is the Hamiltonian in black hole units.
According to thermodynamics, the thermal density matrix is \(\ex{-\bet H}\), where \(\bet=1/T\), so this gives the temperature \(T=1/\pi\). 

Hawking did the calculation in a slightly different way, comparing the black hole with Rindler space.
The local vacuum state, \(|\W\ket\), may contain \(n_I\) particles in region \(I\) and \(n_{II}\) particles in region \(II\).
He calculated the probability that there are \(n\) particles visible in region \(I\):

\be P(\, n_I=n\,)
=C^{st}|\sum_{n_{II}}\bra \,n_I=n,\,n_{II}|\W\ket|^2=C^{st}|\ex{-\pi n\w}|^2=C^{st}\ex{-2\pi H}\ .
\nm\ee
Therefore, he found a temperature \(1/\bet =1/2\pi\).
This disagreement is important. Forcing the number of particles in region \(II\) to be equal to the number of particles in \(I\) was the primary reason for this difference, but it also is the reason why the Fourier transforms mentioned above are unitary in the subspace \(I\) only. Therefore, if we want a theory where the quantum information is preserved, we should accept this new, higher value for the temperature of a black hole.	

The value for Rindler space stays what it always has been. Our theory simply states that black holes have nothing to do with Rindler universes.

The next section contains 11 subsections where we repeat the arguments and calculations in a more elaborate way,  but still without too technical derivations.
	 \end{quote}
 \tableofcontents
\newsecl{Introduction to this chapter}{intro.sec} 
\subsection{The meaning of quantum logic\labell{qu.sub}}
When this book was written, Quantum Mechanics was approximately one century old, and during all this time this theory was presented as a marvel of 20\th century analysis, requiring an entirely different way of thinking not only about space, time and matter, but also about \emph{reality}. This torment of logical thinking was almost thought to be over now; the non existence of local reality was accepted and a lesson was learned: 19\th  century logic was over, whereas surprises would continue to come.

This left the author of this chapter as a solitary bastion, being one of the last representatives of 19\th century logic. He still maintains that classical logic is not over, but our understanding of the equations is still wanting. In the 1960s, our attitude turned out to pay off in a big way, when  new pieces of insight told us that elementary particles can be tamed with equations from \emph{quantum field theory}. We learned to ignore warnings from science philosophers who claimed that special relativity is incompatible with quantum mechanics and locality. What we  did learn was that we should rely on the results of experimental scientists, whose measurements indicated that fields are important. 

Imposing internal logic there, led to light-cone commutation relations and dispersion equations. The procedure of \emph{renormalisation} was found to be an essential ingredient of these theories, and the scheme arrived at became known as the \emph{Standard Model of the elementary particles.}  Thanks are due to numerous experimental physicists who made   relentless efforts to figure  out how to measure almost everything there was to know about elementary particles. It was  concluded that quantum field theories are just fine in spite of what those philosophers said. For ordinary elementary particles, the right equations were discovered, and how to use these was almost completely understood. These particles obey quantum mechanics as well as special relativity -- \emph{and} locality.

My claim\,\cite{fastvar.ref}%2
\cite{GtHHarmont.ref} %3
is that the world can be described by local non-quantum equations, and  classical logic will be fine. According to this theory, the reason why quantum computers will seem to do calculations of unbelievable complexity and rapidity, is that they will be able to  harvest classical vacuum fluctuations that will go all the way to the Planck scale. Yet unlike standard quantum mechanics, these fluctuations may represent only one reality.

This book is not about the interpretation of quantum mechanics. It is about including \emph{general relativity} in our quantum models. And now the new problem is that experimental information about this question is very difficult to produce. Our present strategy is to apply quantum mechanics to black holes. Our hope is that, just requiring that the answers should be  sensible, could yield new information on how to proceed. But, as one will see, the math is quite demanding.

This chapter is organised as follows. The remainder of this Section \ref{intro.sec} contains most of our arguments.  first, we show how Hawking's result can be understood as a consequence of Bogoljubov transformations acting on the region surrounding the horizon.
Our first attempt leads to a picture of particles radiating away from a black hole,
 which we call `scenario 1'.  This scenario assumes that black hole radiation has precisely the temperature as calculated by Hawking\,\cite{Hawking.ref}% 4
.  But then we show that the Bogoljubov transformation can be reconciled with preservation of quantum information, only if we use `scenario 2' for the interpretation. Scenario 2 however would imply that the temperature of the radiation is twice as high as Hawking's value.
 
We  introduce the various techniques that are indispensable to know about, using a sequence of subsections.
During this entire first section, subsections \ref{qu.sub}--\ref{disc.sub}, we shall avoid most of the equations. 
Therefore, this part of the chapter is not too technical. In the sections \ref{SScoor.sec}--\ref{infotemp.sec} we display the calculations backing our result.

\subsection{Quantum mechanics in black holes}
Black holes occur at all scales in physics, and therefore it is conceivable that their role in uniting general relativity with quantum mechanics may well turn out to be crucial. 
We  use the quantum language here the way we were taught to, as it is described in our elementary text books; just don't leave reality out of sight. In particular, resist the temptation to make our cover story about quantum mechanics look even more fantastic than it does now. Don't let observers run through wormholes and the like. Our first attempts must be to bring our understanding of the gravitational force to the same level as our understanding of elementary particles. 

The primary reason why the basic synthesis of gravity with quantum mechanics is so difficult to understand is, that direct experimental data are so much more difficult to acquire.  As could be expected, a majority of researchers in this field concluded that black hole theories would have to be non-local\,\cite{Giddings.ref}. %5
I do not believe them, and there are no non-local phenomena to be found in the treatment proposed in this chapter, but we are not there yet;  let us continue. 

Throughout this chapter, our basic goal is to arrive at unitary evolution equations. To get an impression, the reader is invited to glance through the last pages (such as Sections \ref{clones.sec} and \ref{infotemp.sec}), just to see in what direction we are going. We do not succeed completely, but we think we are on our way.

\subsection{Schwarzschild coordinates} We introduce the Schwarzschild coordinates in Section~\ref{SScoor.sec}. At first sight, they do not reveal much about the numerous excited states a black hole can be in, but we can easily formulate these excited states in terms of particles that may be frolicking around near the horizon. The case we are primarily interested in is the black hole that is large and heavy compared to the Planck scale. One then finds that the particles that give more structure to the horizons, will  be much lighter than the Planck scale. This allows us to treat the Schwarzschild metric as a sturdy background, and the equations for the light and tiny particles can be derived from the Standard Model, generalised to handle the case of weakly interacting particles, in a moderately perturbed background metric.
\subsection{The no hair theorem}
Note that we ignore the `no hair theorem'. This theorem is one of the errors one can commit by not understanding 
 the laws already known. Dust falling into a black hole has its time reverse coming out; together they form lots of hair. Time reversal symmetry applies to all of the known laws of nature at small scales, so why forgetting that?
 
\subsection{The Kruskal-Szekeres coordinates} The weakly perturbed black hole can be analytically extended to form a maximal analytic structure, using the Kruskal-Szekeres coordinate system, to be introduced in Section~\ref{KS.sec}. This description of the same space-time for which the Schwarzschild coordinates had been introduced, is nevertheless quite different: it connects the universe surrounding the black hole to another universe, whose role will at first be mysterious. It is necessary first to include this `spectator universe' in our descriptions, to leave its true role in the physics of black holes for later. Only near the end, in Section~\ref{infotemp.sec}, the building blocks of our theory will merge into a solid picture. It has to do with conical singularities. 

\subsection{Vacuum states} The vacuum state is defined to be the state from which no particles can be removed. That is, the annihilation operators \(a\), to be derived from the fields and their equations, just give us the null state:
\( a|\psi\ket=0\ .\)
 But wait, one can always consider the removal of particles that were involved in `dressing' the black hole, and indeed, technically, this means that genuine black holes cannot be surrounded by a real vacuum. Or in other words, there is only one vacuum state, which is the state where we removed the black hole itself. 

But we can reformulate empty states in the immediate vicinity of a black hole. These are good approximations for describing a black hole with no other particles in their vicinity, and their role can be derived accurately: there are different kinds of such vacuum states!  In Sections 4 and 5, it is explained why this definition of vacuum states depends on who is watching. The observer outside the black hole or someone very close to the horizon? It depends on the time coordinate used by the observer.

\subsection{Thermal radiation emitted by a black hole: Scenario 1} In Sections~\ref{globallocal.sec} and \ref{bogoly.sec}, we shall handle the transformation from one time coordinate to another, and explain the resulting Bogoljubov transformations, which do not keep the vacuum states invariant.

Next, in Sect.~\ref{Hawk.sec}, we describe how Hawking found that a distant observer sees  a black hole that emits radiation. This radiation is purely thermal, and he finds a  temperature 
\be T_H\qu 1/8\pi k GM\ . \eel{THawkqu.eq}
In this section, we derive how our formalism can yield this result, including this value for the temperature, by assuming a prescription that we call `scenario 1'. 

\subsection{The thermal density matrix: Scenario 2} 
%\vskip200pt
The above calculation was borrowed from a similar calculation in \emph{Rindler space}\,\cite{Rindler.ref}. %6
Yet there is something odd with this result (hence the question mark in Eq.~\eqn{THawkqu.eq}). It was derived from the wave function of the vacuum state \(|\W\ket\) for the local observers. In terms of the particles in regions \(I\) and \(II\), this state is found to be 

\be\Psi=C^{st}\sum_{n_{II}\ge 0} |n_I\ket\, \ex{-n_I\pi\w}\,|n_{II}=n_I\ket\eel{psin1=n2.eq}
(multiplied over all positive values of the variable \(\w\)), and in view of the restriction \(n_{II}=n_I\)\,,
there is hardly any summation.
However, if it were really a thermal expression valid for two independent universes, then all states with arbitrary values for \(n_I\) and \(n_{II}\)  should be included. The fact that, in the expressions we get, \(n_I=n_{II}\), implies that region \(II\) is just an exact copy of region \(I\). That is not quite the situation in Rindler space. In Rindler space, one really has two different universes connected naturally at one point (more precisely: a sheet, if we add the transverse coodinates).

The expression obtained from the Bogoljubov transformation is actually an amplitude, to be squared to obtain the probabilities of the various states in region \(I\). 

This is also the reason why, in Rindler space, one can never recover information that disappeared across the horizon. If, in contrast,  we insist to describe a black hole in such a way that information is returned, we have to identify the two universes as being a single one.

\subsection{The quantum clone} 

This is why we advocate to use not scenario 1 for a black hole, but scenario 2, which states that the wave function we find by employing the Bogoljubov transformation, actually plays the role of a \emph{density matrix} for the black hole:
\be \eqn{psin1=n2.eq}\ \ra \  \rr=\ex{-\bet H}\ ,\eel{densmatr.eq} for a real thermal density matrix, but this differs in various ways from scenario 1:

First, a density matrix naturally only consists of bras and kets describing exactly the same physical state, which explains our restriction \(n_I=n_{II}\).

Second, 
the bra state \(\bra\psi|\) runs backwards in time. This is why it is actually controlled by \emph{minus} the Hamiltonian for space \(I\).
Consequently, this density matrix would apply only for one universe. Being a density matrix, it should not be squared to obtain the probabilities -- it is a square already. 

But this implies that, in scenario 2, the exponent being \(\ex{-\bet E}\),   gives a  \(\bet\) whose value is half of what Hawking found; so the temperature is now twice his.

Thus one finds a temperature twice the values given by Eq.~{THawkqu.eq}:
\be T\overset{\textstyle !}{=}		1/4\pi k GM\ . \eel{T-GtH.eq}  

In Section~\ref{regions.sec} we discuss this difference. The question that must be considered concerns the interpretation of the universe in region \(II\) of the KS metric. Do we have indeed an other universe there? We claim to have convincing arguments that region \(II\) is not another, `alien'  universe. It is a copy of our own universe. We explain this in Section~\ref{regions.sec}, and add more clues in Section~\ref{clones.sec}.
For black holes, scenario 2 is indeed more natural than scenario 1.

\subsection{The Shapiro effect} A crucial ingredient in our treatment is the requirement that a radiating black hole should behave in agreement with the fundamental laws of quantum mechanics:  the Schr\"odinger equation can be inverted in time, so that, consequently quantum information should not get lost. A beautiful mechanism for the retrieval of black hole information is obtained by considering the gravitational forces produced by the Shapiro effect. This is the gravitational pull by which particles going into a black hole affect the particles coming out. The emerging particles are shifted along the horizon where they originate. How this effect can repair information loss is discussed in section \ref{infotemp.sec}.

\subsection{Discussion\labell{disc.sub}}
This work is not finished. There are several weak points in our approach. We were forced to use approximations and simplifications. For instance the Shapiro effect that we use to see information be returned to us in the form of Hawking radiation, only  partly does its job: only \emph{geometric features} generate the gravitational forces that send particles back out of the black hole. It remains to be seen whether among the `in particles', quantum numbers not related to energy and momentum can be restored to re-emerge with the `out particles'. 

And then, consider the crossing points of future and past horizons. The author has neither time nor space to do further calculations on this singular point; it might well turn out to play a pivotal role in future formalisms. Let us just observe that, identifying regions \(I\) and \(II\), imply a division of space-time by an isotropy operator \(\ZZ(2)\). It actually acts as a mirror, the ideal instrument to send information back to where it came from, but also it induces a mild \emph{conical singularity} This singularity actually has some of the characteristics of a membrane. One can imagine membranes of this sort to affect the Standard Model interactions, but how this goes is not understood at present.

These are not drawbacks that would be likely to invalidate our ideas; rather, they pose new challenges, to be addressed.

In any case, our procedure works best if the two universes are handled as being exact \emph{quantum clones} of one another. That doubles the temperature of the radiation. And we resist the temptation to return to the subject by which we began writing this chapter: It may well be that, eventually, the interpretation of quantum mechanics will have to enter this discussion. 

The rest of this chapter, subsections \ref{SScoor.sec} - \ref{infotemp.sec}, handles some of the technical calculations.

\newsecl{Schwarzschild coordinates and horizons}{SScoor.sec}

\subsection{The region \(r\ra 2GM\)\labell{green.sec}}
	Shortly after Albert Einstein wrote his famous equations for the gravitational force, well-known astronomer Karl Schwarzschild\cite{Schw.ref} found an exact solution for the spherically symmetric case, in the absence of matter:
\be \dd s^2=\Big(\frac{2GM}r-1\Big)\dd t^2+\Big(1-\frac{2GM}r\Big)^{-1}\dd r^2+r^2\dd\W^2\ ; \quad\Big\{ \  \begin{matrix} 
\W&\equiv&(\tht,\vv)\, ,\\[3pt] \dd\W&\equiv&(\dd\tht,\,\sin\tht\,\dd\vv)\,,\end{matrix}\quad  \eel{Schwmetric.eq}
where \(\dd s\), the distance between two infinitesimally close space-time points \(x\) and \(x+\dd x\), is defined to be real in the space-like direction and imaginary when time-like.
This metric describes a black hole of mass \(M\) that remains unchanged both in the future and in the past. 

Its space-time geometry is sketched in Fig.~\ref{SchwMetric.fig}a. The angular coordinates \(\W\) are suppressed here; the dashed vertical line is the \emph{horizon}, corresponding to the point  \(r=2GM\). We see that the coefficient im front of \(\dd t^2\) vanishes there, while the coefficient in front of \(\dd r^2\) generates a singularity. This singularity can easily be removed by replacing the radial coordinate \(r\) with \(\rho\)\,:
\be \rho=\sqrt{r-2GM}\,,\quad\hbox{so that}\quad r=2GM+\rho^2\ ,\quad\hbox{and}\quad    
\dd r=2\rho\dd\rho\ .\eel{rhopar.eq} In terms of these coordinates, the metric \eqn{Schwmetric.eq}
can be written in a compact form: 
\be \dd s^2=\frac{-\rho^2\,\dd t^2}{r^2}+4r\,\dd\rho^2+r^2\,\dd\W^2\ .
\eel{rhometric.eq}

Note that Eq.~\eqn{rhopar.eq} implies that \(\rho\) is imaginary when \(r<2GM\). This will make the region \(r<2GM\) unphysical to some extent -- but not quite: it is better to say that the radial space coordinate becomes time-like and the time coordinate becomes space-like there,
see the blue shaded regions in Figs.~\ref{SchwMetric.fig} and \ref{SS-KS.fig}. Later we will see more precisely what this means, and indeed the region \( r < 2GM \) will only play a secondary role.

\begin{figure}[h!] \qquad
 a)\widthfig{200pt}{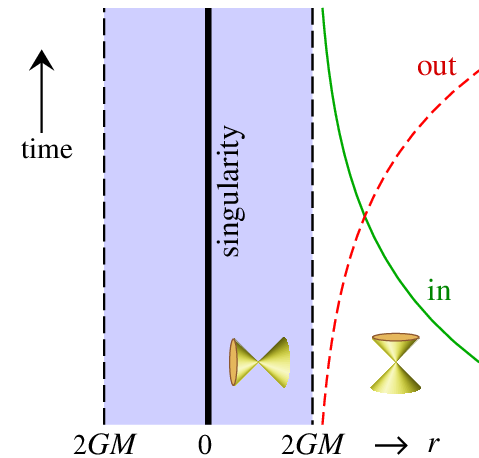}\qquad\qquad b) \lowerheightfig{10pt}{220pt}{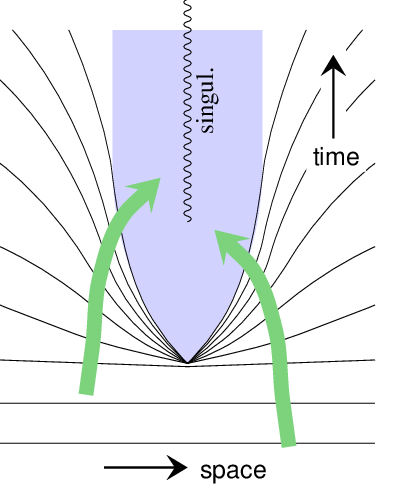}
\caption{\small a) The stationary black hole, in the Schwarzschild  coordinates,\((r,\,t)\),  showing the geodesics of a particle going straight in (green) and a particle going straight out (red, dashed).  
The dashed vertical line  is the horizon; the cones indicate the orientation of the local light cones. The angles \((\tht,\,\vv)\) are not shown. The central singularity cannot be crossed. \newline
  b) An artist's impression of spacetime for a black hole being formed. The black curves are constant \(t\) lines, approaching the horizon. Thick green arrows represent  imploding matter. The wavy line is the central singularity.
\labell{SchwMetric.fig}} \end{figure}

Note on the other hand, that this metric naturally continues to \emph{negative} values of \(\rho\), where, as we shall see in the next section (or by carefully examining Eq.~\eqn{rhometric.eq}),  the physics is exactly the same as for the positive values. Nothing becomes complex or imaginary there. However, clearly, this seems to make   space-time twice as big as what it should be. It looks as if black holes are some kind of worm holes towards another universe, but it is a big mistake to think that. We return to this feature when the regions \(I\) and \(II\) of the Kruskal-Szekeres coordinates are discussed, see section \ref{KS.sec} of this chapter.

\subsection{The time coordinate \(t\)}

Allow us to make a somewhat extensive digression. For those who prefer their tea stirred anti-clockwise: we cannot proceed without observing that the black hole's past seems to be represented incorrectly in the metric as represented in Fig.~\ref{SchwMetric.fig}a. This is often brought forward as an objection against using the metric as represented in Eq.~\eqn{Schwmetric.eq}. And it is true that all matter that has entered into the black hole during its entire life in the past, should still be visible at the horizon, the vertical dotted line. The collection of all in-going particles (in-particles for short) as represented by all green solid curves below the curve shown, must have an effect on the horizon. In fact, they change the entire metric into the one sketched in Fig.~\ref{SchwMetric.fig}b. Therefore, the approach starting with the eternal (that is, time-independent) metric \ref{SchwMetric.fig}a is often dismissed.  Authors then include the history as they prefer to see it, which includes the event that created the black hole, no matter how long ago that was. For simplicity, one often imagines a spherically symmetric cloud of matter, which collapsed under its own weight.

However, in doing so, one would completely ignore the fact that both quantum mechanics and Einstein's equations for General Relativity,  are  invariant not only under time translations but also under time-reversal. The standard arguments in the literature\,\cite{Giddings.ref} rarely take into account that the black hole, when it  eventually evaporates completely, should be handled as the time-reverse of the implosion that created the black hole in the past\,\fn{S.W.~Hawking\,\cite{Hawking.ref}\cite{Hawkinginfo.ref}\cite{Hawkdens.ref}  
 did take into account that the black hole may evaporate completely, but the space-time metric he produced was asymmetric under time reversal. This was because he took into account the fact that the \emph{entropy} of any configuration will be allowed only to increase, and that would be asymmetric under time reversal. However entropy refers to the probability density of states, not the states themselves, which should display invariance or covariance under time reversal, or CPT symmetry to be precise, as we know from elementary particle physics.}.

The evaporating particles (out-particles), represented by the red dashed line in Fig.~\ref{SchwMetric.fig}a, should be treated just like the in-particles are. All evaporating particles in the future cause similar deformations of the metric near the horizon. In this entire chapter, we shall treat the features of the collapse and of the evaporation in a time-symmetric way.
 \subsection{The heat bath\labell{heatbath.sub}}
The way to handle the situation systematically is now as follows: \\
we assemble all possible states of a black hole by first considering how these states should behave when we assume  the black hole to be stationary. This is realised if the black hole is situated in a heat bath formed by particles that have the same temperature as the Hawking particles generated by the black hole.
Then, consider the region near the center where all in- and out-going particles meet. Let us assume that an observer there sees no particles. In that case, the situation should be completely stationary. As we shall derive however, if a local observer sees no particles, not only does a distant observer in the future see  Hawking particles emerging from the hole, but by time reversal invariance, an observer in the distant past should also see particles entering the black hole -- that is, the particles in the heat bath. In conclusion, the distant observers might not agree with the local observer concerning the particle density, but they both agree that the situation can be looked upon not only as entirely stationary in time, but also symmetric under time reversal.\fn{Though stationary, this black hole will not be absolutely stable; this is not  a  problem in practice, as long as we restrict ourselves to describing relatively short stretches of time.}\ret
Time reversal invariance is then to be regarded as a necessary property of \emph{all possible microscopic laws }of physical evolution, where entropy does not yet play a role; this is a feature that occurs in many branches of physics.

\subsection{Small excitations\labell{smallex.sub}}
	To consider more complete sets of black hole solutions, we subsequently add or remove particles by applying creation and/or annihilation operators defined by the ubiquitous elementary particle fields. This then allows us to consider black holes with time-dependent features.
One may always consider entangled particles later; they will not affect the primary equations affecting the original Schwarzschild solution.

	This is the way we go in this chapter. Our point is that both the distant past and the distant future will be irrelevant as soon as one can describe the evolution of the entire set of \emph{all states that may be involved} and that we may encounter  in any given short stretch of time (see the green region in Fig. \ref{SS-KS.fig}a). Every element of this set evolves into an other element of this set. The set should be defined in such a way that neither the states describing the distant past, nor the ones in the distant future, have a direct effect on the evolution law in the present. Think of an electron. The equations from which one can derive its behavior at present, involving the entire Hilbert space, do not require the knowledge whether the electron emerged long ago by pair creation, or beta radiation or whatever. Neither do we need to know how it will end its life.
	
	Important remark alluded to earlier: 	at times earlier or later than the green region, the particle excitations eventually will also invade the blue regions but we shall have to ensure that all important events that will give us local physics will be at the visible sides of the black hole only (or possibly we'll enter with infinitesimal steps in the forbidden domains). Thus, there is no need to worry about particles moving from the green zone to the forbidden region close to the horizon; they are taken care of by the mechanism alluded to earlier and to be discussed later in this Chapter.
	
We only put particles where they can directly be seen, or at least their interactions must play a role \emph{at the present}. This means that, for the time being, we \emph{don't} put particles anywhere in the blue regions of Figures~\ref{SchwMetric.fig}a and b, or Figures~\ref{SS-KS.fig}a and b. What happens in those regions is postponed to later concern (Section \ref{KS.sec} and onwards). This is the first and not at all the last remark concerning locality, an important concept in this approach. So far, we have merely added small particles in the green region, at some safe distance from the horizon; how these evolve in earlier or later domains remains to be calculated.
	
	The state we choose for a black hole, at any given time, which may include entanglements of several states, depends on the past. But in general, the physical laws here are well captured by the Schr\"odinger equation active in the present, just as we have it for all particles in the Standard Model, where any further reminders of the past are unnecessary. This simple observation already departs from claims by some other observers, which we ignore until forced otherwise (and this does not seem to happen anywhere in the description we shall arrive at, in this chapter). 

	Eventually, of course, we should formulate the entire set of equations to see what really happens. This is our aim. Thus we proceed using the metric of Figure \ref{SchwMetric.fig}a. It is used as a starting point: neither the distant past not the distant future can affect the e.o.m. for the behaviour of the black hole at present. But, as we shall see, the complete set of all excited states of a black hole, having light particles flying around everywhere in the green region, needs to be defined with care. More about this in Section~\ref{shapiro.sec} of this chapter; we shall indeed find that one can define other excitations, but, surprisingly perhaps, these can be transformed to elements of the set we already have.

	End of digression -- although this needs not be the end of the discussion of time reversal symmetry. We are aware of the fact that nature does not \emph{have to} be symmetric under any symmetry transformation of the form of CPT. We find that our first attempts to formulate a synthesis of general coordinate invariance and quantum mechanics must share the known symmetries of both theories. It is entirely acceptable to study other options, but we leave that to other investigators. 
\newsecl{The Kruskal-Szekeres coordinate frame}{KS.sec}
\subsection{Using short -- but not too short -- time intervals}
The conclusion of the previous subsection is that we should first try to use the metric of a completely stationary black hole, and consider particles added to that (or removed, see later), in order to get the most generic states.
Starting from that, we plan to construct the Schr\"odinger equation\fn{in the not so distant future, but not in this chapter, unfortunately.}, which must be accurate during the entire evolution process of the black hole. First we must make sure that the equations work accurately during a sufficiently short time interval. 

During such an interval, the black hole mass will stay almost constant, but one can be more precise, by allowing the mass to make small, perturbative fluctuations.\fn{It is easy to imagine where fluctuations in the mass can come from: particles will be flying in and out. The mass/energy of the entire system is constant, but the mass of the black hole itself  depends on the extent to which we regard particles very close to its horizon, to be integral parts of the black hole, or alternatively,  whether a particle may be considered to be separated too far for that.} Since the mass is observable, standard quantum mechanical techniques will suffice to understand how the mass will slowly evolve during longer time intervals. This will in practice be exactly what we need to know. 

Of course, an important question will be how to knit successive time intervals together; we'll come to that.
Having that, the next task will be to see how transitions to lighter or heavier black hole states can take place, and this should enable us to handle the initial and final states of the black hole correctly after all.

Space-time will look as in Fig.~\ref{SS-KS.fig}a. Here, the green part is the region that we allow to be filled with particles of the Standard Model (or beyond). All particles must be sufficiently light to allow us to neglect their gravitational fields, \emph{in as far as they stay in the green region}. 
However, their geodesics will continue, and either in the far past, or in the far future, or both, they will approach a horizon too tightly to be called a particle at all. As seen by the local observer, heir energy and momentum will increase exponentially as we get further away from the green region. It so happens that we have identified an \emph{other way} to represent such an `ex particle' as an elementary component of the black hole state. This procedure will remove exactly all particles that we do not want to regard as particles anymore. This is further explained in Subsection~\eqn{timeslices.sub}.

% and we shall discuss the beutiful ensuing physics later in this chapter.

\begin{figure}[h!] \bcenter   \widthfig{540pt}{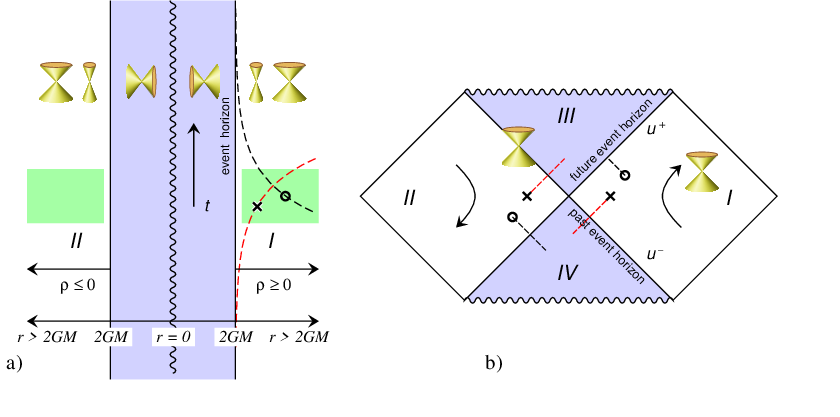}
	\caption{a) \small Schwarzschild metric showing coordinates \((\rho,\,t)\). Solid angle \(\W\) is suppressed. Green is the time slot where we choose the in- and out-going particles (marked O and X). The yellow cones show the orientation of the local light cones in the various domains.\newline b) Kruskal-Szekeres\,\cite{Kruskal.ref}%8
\cite{Szekeres.ref} %9
coordinates in the form of a Penrose diagram. The coordinates \(u^\pm\) are defined in Eq.~\eqn{upmdef.eq}.    Region \(I\) represents the outside region in (a), region \(II\) will first be treated as if the physical events taking place there are independent of the ones in region \(I\).  Only later we shall consider the option of quantum clones. Arrows in (b) show how local time proceeds. In both figures, the yellow cones show the   orientation of the local light cones. % check ... where?
 \labell{SS-KS.fig}}\ecenter \end{figure}
\subsection{The coordinates \(x\) and \(y\)}
We have now come to the point of considering other coordinate frames for the same metric, that are not singular on the horizons (while the original Schwarzschild coordinates do generate singularities there). Nevertheless, we need to keep track of time translation invariance. The ideal coordinate system for this purpose\fn{\,There are other choices that can be useful but they will not be considered here.}  turns out to be the Kruskal-Szekeres 
coordinates\,\cite{Kruskal.ref}\,\cite{Szekeres.ref}, where  \(r\) and \(t\) are replaced by \(x\) and \(y\). These are defined by
\be x\,y  &=& \big(\frac r{2GM}-1\big)\ex{r/2GM}\ ,\nm \\[5pt]
y/x &=&  \ex{t/2GM}\ .\qquad \eel{KS.eq}
One derives that in he Kruskal-Szekeres frame, the metric \eqn{Schwmetric.eq} reads
\be \dd s^2 &=& \frac{32(GM)^3}r\,\ex{-r/2GM}\dd x\,\dd y\ +\ r^2\dd\W^2\ .  \eel{KSmetric.eq}

For later use, we introduce a new dimensionless time parameter \(\tau\),
	\be \tau&=&t/4GM\ , \quad\hbox{so that}\nm\\
	y&=&\ex{ q+\tau}\ ,\qquad x \ =\ \ex{q-\tau}\,,\quad \hbox{where} \nm \\
	q&=&\half\log\big(r/2GM-1\big)+r/4GM\ .
	\eel{taudef.eq}

% check nog voor later gebruik
An important feature of this re-parametrisation is that it is not 1-\,to\,-1. Every point  \((r,t,\W)\) in the Schwarzschild frame, Fig.~\ref{SS-KS.fig}a, is mapped both to \((x,y,\W)\) and to \((-x,-y,\W)\). Indeed this repetition is fundamentally \emph{exact}, a feature that is often ignored. Does any event  in region \(I\) have a \emph{quantum clone} in region \(II\)? Does an observer have a quantum clone? We shall return to this (Sections \ref{globallocal.sec} and \ref{regions.sec} of this chapter).	
	%will we? !!

\subsection{Penrose diagrams}
After replacing the light cone coordinates \(x\) and \(y\) by coordinates \(u^\mp\) that  cover a finite domain, one gets a picture of space and time in the form of the  diagram as sketched in Figure \ref{SS-KS.fig}\(b\).  A good choice is
\be  x=C\tan {u^-}\,,\ \ y= 
 C\tan {u^+}\,,  \quad\hbox{ with } \quad
 C=  \frac{\sqrt{e/2}} {2GM}\ ,\quad |u^\pm|<\half\pi\ .
 \eel{upmdef.eq}
The picture of the universe that we thus obtain, has the useful property that local light cones everywhere are oriented in the same direction (see the cones in fig.~\ref{SS-KS.fig}b). Such diagrams are called \emph{Penrose diagrams}\cite{Penrose.ref}%10
\cite{HawkEllis.ref}. %11.

One good thing about these coordinates is that they are {conformal}, which here means that the coefficients in front of the terms with \(\dd x^2\) and \(\dd y^2\) both vanish. A consequence of this is that, at given, fixed angles \(\W\), the lines with \(x\) constant and the lines with \(y\) constant, are light rays. If we rotate the (\(x\,,y\)) frame by \(45^\circ\), as in Fig.~\ref{SS-KS.fig}b,  we see that time-like curves all have a slope of less than \(45^\circ\) with respect to a vertical line, while information cannot travel along the lines with slopes larger than \(45^\circ\). The transformation to the coordinates \(u^\pm\) shown in Eqs.~\eqn{upmdef.eq} was chosen to turn the diagram shown in Fig.~\ref{SS-KS.fig}b into a compact shape. The boundaries of the 
physically important regions are formed by the lines \(|u^\pm|=\half\pi\) and \(u^\pm=0\). The local light cones are  oriented as is shown in the figure.

The only singularities in the coordinates \((x,\,y)\)  (or \(u^\mp\)) are at the edges. The wiggly line is a true natural singularity, but that is entirely in the domain where \(r\ra 0\), so that \(\rho\), \,see Eq.~\eqn{rhopar.eq}, is imaginary. One also notes\fn{This suffices to ensure that these singularities are also physically insignificant. The future event horizon itself is already at the infinite future for any observer in region \(I\). We would like to keep things that way, but remarkably, we may observe later that infinite future and infinite past might become interchangeable.} that this natural singularity is always beyond the infinite future (in region \(III\)) or before the infinite past (in region \(IV\)).  

One can now see that there are two different horizons. The solid line where \(u^-=0\) is the \emph{future event horizon} and the line \(u^+=0\) is the past event horizon.

In particular, the metric is completely smooth at the origin of the diagram. There is, however a very peculiar, somewhat worrying aspect of this particular Penrose diagram: it seems to consist of \emph{two universes}, connected at the origin. Only space-like curves rigorously connect these two universes. What does the second universe stand for? Many authors  have different feelings about what it means for a black hole to involve two universes. How do we understand this physically? One popular view is what here will be called the `science fiction' view: a traveller who, somehow, manages to move faster than light during a short lapse of time (\emph{eigen} time to be precise) should be able to travel from one universe to the other. Unfortunately, the present author does not believe in such science fiction stories (one can't travel faster than light), so this option will quickly be dismissed. 

We shall frequently refer to the different regions in the  Kruskal-Szekeres (KS) coordinate frame\, 
\cite{Kruskal.ref}\cite{Szekeres.ref}, which we label \(I,\ II,\ III\,,\) and \(IV\), as shown in the diagrams.\fn{Other authors favour the interchange \(II\leftrightarrow III\), so that their labels are ordered anticlockwise. We prefer the order used here, as regions \(III\)  and \(IV\) often play not such a big role in our derivations. As yet, particles in regions \(III\) or \(IV\) seem to be insignificant -- but please hold on.}
And indeed, fortunately, there are more rigorous procedures to figure out how to deal with the physics of what we call region \(II\), as we shall see.

Let us emphasise again: region \(II\) appears to be an exact copy of region \(I\). 
In figure ~\ref{KSCauchy.fig}a, we see how other researchers often describe the time evolution in the Kruskal - Szekeres coordinates. they take this diagram as an ordinary manifold where region \(I\) describes our entire universe, while region \(II\), an exact image of region~\(I\,\), somehow stands for something else, \emph{inside} the black hole. To us, region \(II\) cannot be anything else than a copy of our own universe. It only makes sense to assume that it is an exact copy of our universe, a quantum clone, to be precise, but we shall come to that, see Sections [\ref{regions.sec}] and [\ref{clones.sec}].

\subsection{Cauchy Surfaces}

The time evolution of physical states (states with particles flying around in regions \(I\) -- \(IV\)) is to be described by drawing different \emph{Cauchy surfaces},   3-dimensional subspaces of the space-time manifold  where we describe all possible states at a given moment of time. 

In flat space-time we can consider all planes where \(t=C\)\st\ to form a useful set of Cauchy surfaces enumerated by this constant. 

For a black hole described by the KS metric we  use the  \(\tau=0\) plane (see Eqs.~\eqn{taudef.eq}) that separates the \emph{future} from the \emph{past.}  This is our primary Cauchy surface. As time  \(\tau\) proceeds, many researchers draw the Cauchy surfaces as in Fig.~\ref{KSCauchy.fig}a. This would be the picture drawn by a local observer. However,  as time \(\tau\) proceeds, time in region \(II\), as defined by a distant observer, runs backwards. Therefore, one should draw the Cauchy surfaces as in Fig.~\ref{KSCauchy.fig}b. These  are the surfaces \(\tau=C\)\st. 
Indeed, he most remarkable feature of this metric is that in one part of  spacetime, region~\(I\), time \(\tau\) runs forward while, equally swiftly, time in the opposite part, region~\(II\),  \emph{runs backwards}, see the arrows in this Figure.

\begin{figure} 
a)\hskip-15pt\lowerwidthfig{10pt}{240pt}{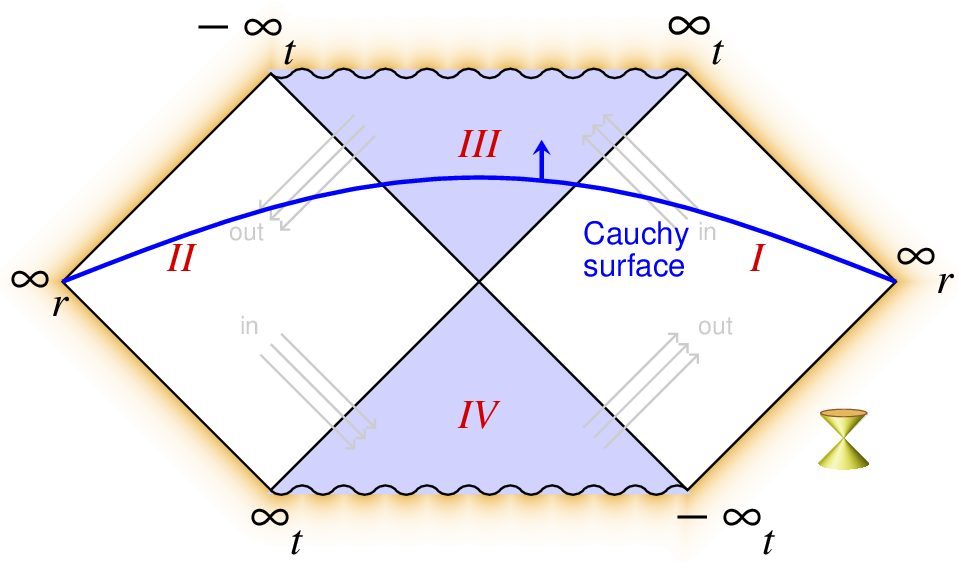}\qquad 
\ b)\hskip-20pt\lowerwidthfig{10pt}{240pt}{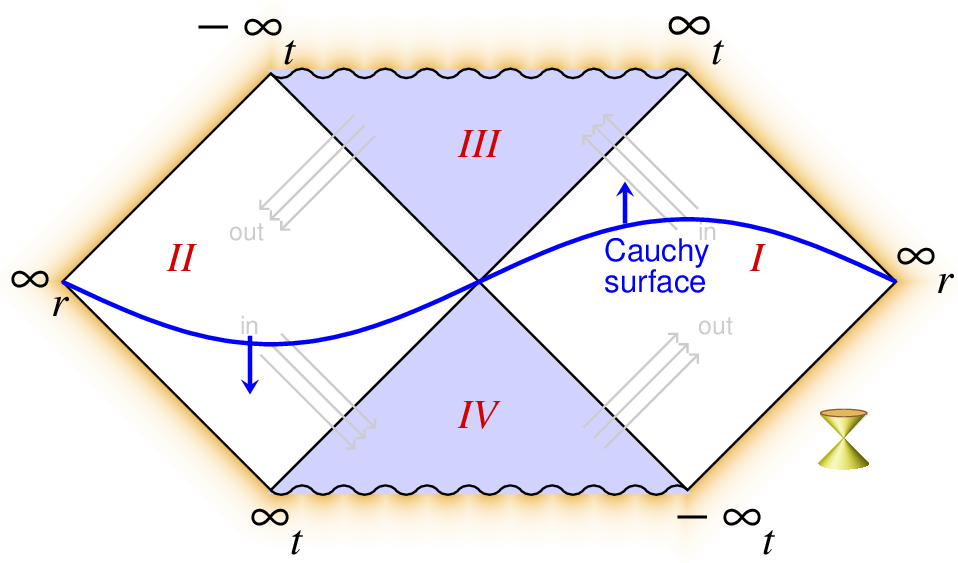}
\caption{a) \small Cauchy surface (blue curve) as often adopted in the literature. This choice is not compatible with time translation covariance. b) Cauchy surfaces of this shape form a set that is completely invariant under time translations and therefore our favoured choice.\labell{KSCauchy.fig}}
\end{figure}

The novelty of this situation is that, the point at the center stays fixed to the center at all times as seen by the distant observer. It so happens that the blue line in Figure~\ref{KSCauchy.fig}b is much more suitable than Fig.~\ref{KSCauchy.fig}a\, for the calculations to be done next. It represents  a valid set of Cauchy surfaces, since the data at \(\tau=0\) completely fix the state the black hole is in; at every other fixed value of the  \(\tau\) variable these data can also be fixed.

In regions \(III\) and \(IV\), the direction of the time flow is ambiguous; both these regions may be thought of as representing time, either after the infinite future, or before the infinite past. The data in these two regions are not needed for understanding observed phenomena.

Consequences of this situation often seem to be ignored, while, actually, they are essential. Some authors complained that our choice for the Cauchy surfaces looks too complicated. That's a matter of taste; our essential point is that time translations for the time coordinate, \(\tau=t/4GM\) in Eq.~\eqn{KSmetric.eq}, keep this set of Cauchy surfaces unchanged, a must if we wish to understand time translation invariance (or covariance).

Let us again make an attempt to describe the complete set of states for a black hole, taking into account the fact that we have the two regions, \(I\) and \(II\). We introduce them by first considering the entire black hole region, and, just like most other authors, we consider pumping in particles, one by one, anywhere in the blank regions of Figure \ref{SS-KS.fig}b. For the time being, we allow any kind of particles, both gravitons and Standard Model particles,

 As already stated, this will be an important restriction that we may have to reconsider later, but for the time being it is the first thing to try\fn{Indeed, a suggestion made recently is that \emph{heavy objects}, such as rocks, planets, neutron stars, or even other black holes, may enter into a black hole. It could be that the final states then not only contain familiar clouds of Hawking particles, in the form of gases of Standard Model particles, but also heavier objects, including microscopic black holes. This is an exciting idea that needs to be followed up.}.

This is done often in physics: particles whose interactions are sufficiently weak are represented by fields that only interact linearly. It will be an inevitable simplification, as yet.

We now define particles in terms of  creation and annihilation operators of Standard Model fields, adapted to this curved space-time background. The metric \eqn{Schwmetric.eq} describes the eternal black hole. See Fig.~\ref{SS-KS.fig}a. Only one horizon is visible there, but in terms of the Kruskal-Szekeres (KS) coordinates, we see that there are two horizons, see Fig.~\ref{SS-KS.fig}b. 

It is important to realise how different observers will describe what they see.  Local observers will see nothing special when a particle crosses a horizon. But for a distant observer a horizon is a boundary that cannot be crossed. This observer will see the particle slowing down, become infinitely heavy, and stop completely. It will also become more and more difficult to see the particle. Nevertheless, something changes in the black hole whose horizon it is. We shall see in Section \ref{shapiro.sec}) what happens: due to the Hawking effect, particles come out of the black hole, and these can be affected sufficiently to preserve all information carried in by the in-particles.

The out-particles must have been invisible at early times, but they turn into visible objects later. Naturally, we must understand how to relate those particles that have entered at early times (the early in-particles), to the Hawking particles (out-particles) that emerge at much later times. Can we \emph{restore unitarity} by transforming  in-particles into out-particles and back? In principle, indeed we can! In fact, as will be shown, the mapping is dictated primarily by known laws of physics. And these are local!\,\fn{But there still are important parts that will require further clarification.} See Sect.~\ref{shapiro.sec}.

Thus, on the future event horizon we expect a boundary condition that maps all physical data of the particles crossing that horizon, onto particles that, with an appropriate time lapse\,\fn{This time lapse is precisely dictated by the equations.},  emerge from the past event horizon.

It was found to be crucial that all added particles are excitations due to the action of Standard Model fields, with the restriction that:
\begin{quotation} \noindent a small number \(\eps>0\) exists (to be specified later), such that 
these operators are all situated where they are visible to the outside world,  at \(r>2GM_\BH+\eps\)\,,  inside a sufficiently large interval of the time \(t\). All particles found at earlier or later times, or 
the particles that may have crossed a horizon \((r=2GM_\BH\)), can be included by subjecting the given ones to the evolution equations dictated by the Standard Model, taking in mind the horizon boundary conditions (which in this book will be derived only briefly near the end). \end{quotation} % check ?
We leave for later consideration how to fill the remaining spaces with particles that cannot be seen from the outside. In particular, we must postulate what should be done with particles allowed in region \(II\). As stated before, region \(II\) is exactly as large as the entire universe, normally represented by region \(I\) only.

Since we define the particle content of the black hole by limiting ourselves to a small time interval, there is no place for the imploding particles that produced the black hole in the (distant) past, or for the particles involved in the final puff of the black hole. These must be derived at a later stage by solving Schr\"odinger equations. In line with this, as was elaborated in Section~\ref{SScoor.sec},  we only need to consider, as our background, \emph {eternal black holes}, as described by the metric~\eqn{Schwmetric.eq}.
  
\newsecl{The transformation from  global to local observers}{globallocal.sec} 
\subsection{The central region}
	We  continue with the description of the Kruskal-Szekeres metric, Fig~\ref{SS-KS.fig}, but now we concentrate on a small part of the central region, small enough to allow us to ignore local curvature.
Near the origin, where  \(r\ra 2GM\), or \(\rho\ra 0\) (see Eq.~\eqn{rhopar.eq}), this coordinate frame then obtains the metric \eqn{KSmetric.eq} -- \eqn{upmdef.eq}, or \be\dd s^2\ra  2\dd u^-\dd u^+\ +\ (2GM)^2 \dd\W^2\ ;\eel{dupdum.eq}	
the angles \(\W\) are defined as in Eq.~\eqn{Schwmetric.eq}.	

Any given point \((r,t,\W)\) in the Schwarzschild metric does not correspond to a single point \((u^-u^+,\W)\), 
but to two points: \((u^-,u^+,\W)\) and \((-u^-,-u^+,\W)\), since, in Eqs.~\eqn{KS.eq}, the sign of \(x\) cancels out against the sign of \(y\), see Eqs.~\eqn{KS.eq} -- \eqn{upmdef.eq}. No other ambiguities in the mapping are seen to occur,  see Figure \ref{SS-KS.fig}. 

For large black holes (\(2GM\gg 1\)), the direct effects of single particles on the Schwarzschild metric are assumed to be negligible. These effects take place at time scales much longer than the time scales at which the changes take place that we wish to study, particularly since the particles involved are much lighter than the Plank energy.  Also, we expect that there will be perturbation expansions that can yield sufficient accuracies to allow us, eventually, to study long time scale effects as well. This is why the expressions \eqn{upmdef.eq} and \eqn{dupdum.eq} are accurate starting points for calculations on macroscopic black holes.\fn{In many recent publications the idea is circulating  that particle entanglement features\,\cite{Page.ref} %12
will invalidate such a view categorically. We suspect that this is due to miscommunication; the entropy of the particles that might be expected to emerge from region \(II\) does play a role, but is interpreted quite differently here.
 In any case, we decide first to investigate the more conventional, and to my mind more natural, approaches, focussing on elementary basis elements of Hilbert space, rather than superpositions.}

At a small region near both horizons we limit ourselves  to the longitudinal space-time coordinates \((Z,\,T)\).  We may approximate \(\tan u^\pm\ra u^\pm\), and write Eq.~\eqn{dupdum.eq} as 
\be &\dd s^2=\dd Z^2-\dd T^2\ =\ 2\dd u^+\dd u^-\ ,\labell{locflat.eq}&\\ 
&Z=X^3=\fract 1{\sqrt 2}(u^++u^-)\ ,\quad T=\fract 1{\sqrt 2}(u^+-u^-)\ ,& \eel{loctrf.eq}
(the transverse coordinates \(\W\) and angular momenta \((\ell,\,m)\), may be added later). Note that, the metric \eqn{locflat.eq}, \eqn{loctrf.eq} describes flat space-time; this means that we are performing the flat-spacetime approximation in the immediate vicinity of the central point. 

 We rescale the light cone coordinates  \(\rho\) and \(\tau\) by
 \be &\sqrt{\ds\frac {y\vphantom{u^+}}{x}}=\sqrt{\ds\frac {u^+}{u^-}}=\ex{t/4GM}=\ex{\tau}\,,\quad\hbox{and}\quad Z^2-T^2=\rr^2\,, &\labell{rindler.eq}\\[3pt]
 \hbox{or}\qquad& Z=\rr\cosh\tau\ ,\qquad T=\rr\sinh\tau\ ,& \eel{ZTxytau.eq}	
where \(\rr=2\rho\sqrt r\)  \ (see the definition of the coordinate \(\rho\) in Eq. \eqn{rhopar.eq}),
and \(\tau\) is a rescaled version of the time parameter \(t\) for the black hole, as experienced by a distant observer.
	
\subsection{A scalar field in the central region}
We now need to formulate precisely the transformations of fields in the locally flat region near both horizons, to the fields in region \(I\) and \(II\) of the KS coordinates. From a classical viewpoint, these fields are the same, but in quantum mechanics, their structures in terms of  creation and annihilation operators are different.
An observer using the coordinate frame \((\rr,\tau)\),  sees particles that are not the same as the ones seen in the frame \((Z,T)\). The relation is given by Eq.~\eqn{rindler.eq}. The derivation is not always done correctly. Here is how we derive it (Since \(Z\) is the third space coordinate, we write \(\pa/\pa Z=\pa_3\); furthermore, the transverse component of a 4-vector \(k\) is indicated as \(\tl k\)).

In the coordinates \eqn{locflat.eq}, consider a real scalar field \(\phi(Z,T)\) obeying 
\be \frac{\pa^2}{\pa T^2}\phi=(\vec\pa^{\,2}-m^2)\phi=(\pa_3^{\,2}-\mu^2)\phi\,,\qquad \mu=\sqrt{m^2+\tl k^2}\,,\eel{field.eq}	 
and Fourier transform it in the space direction, defining the function \(k^0(k_3)=+\sqrt{k_3^2+\mu^2}\)\,:
\be\phi(Z,T)= \int \frac{\dd k_3}{\sqrt{4\pi k^0}} \big(    \ex{ik_3 Z-ik^0T}a(k_3)+
\ex{-ik_3 Z+ik^0T}a^\dag(k_3)\big)\labell{phi.eq}\ ;\\		%%
\dot \phi(Z,T)= \int \frac{-ik^0\dd k_3}{\sqrt{4\pi k^0(k_3)}} \big(    \ex{ik_3 Z-ik^0T}a(k_3)-
\ex{-ik_3 Z+ik^0T}a^\dag(k_3)\big)\ .
\eel{phidot.eq}
Here and further on, we use the notation \(\pa_T\phi\equiv\dot\phi\equiv\pa\phi/\pa T\). 

At time \(T=0\), the inverse of this transformation is
\be a(k_3)=\frac 1{\sqrt{4\pi\,k^0(\vec k)}}\int\dd Z\,\Big(k^0\phi(Z)+i\dot\phi(Z)\Big)\ex{-ik_3\cdot Z}\ . \eel{annih.eq}
 Here, \(\dot\phi=\pa\phi/\pa t\), and \(k^0(\vec k) =+\sqrt{ k_3^2+\mu^2}\,.\)  The operator  \ \(a^\dag\ \) is obtained by replacing \(i\leftrightarrow -i\) in Eq.~\eqn{annih.eq}. Notice that these expressions are the standard ones for a scalar, quantised field \(\phi\) in a flat region of space-time. The functions \(a\) and \(a^\dag\) are the standard annihilation and creation operators for particles with momentum \(k_3\) in the \(Z\) direction. The transverse components \(\mu\) of the momenta will be negligible as yet, as these are not involved in the transformations that will be considered. 

At time \(T=0\), we have the usual commutation relations:
 \be [\phi(Z),\,\dot\phi(Z')]=
 i\del(Z-Z'), \quad\hbox{while}\quad  [\phi(Z),\,\phi(Z')]=[\dot\phi(Z),\,\dot\phi(Z')]=0\ , \ee 
 from which one derives \be  [a(k_3),\,a(k_3')]=0\,,\quad [a(k_3),\,a^\dag(k_3')]=\del(k_3-k_3')\ ;
 \eel{comrela.eq}
 (the normalisation factors in Eqs.~\eqn{phi.eq}, \eqn{phidot.eq} and \eqn{annih.eq} were chosen so as to arrive at the usual commutation relations for \(a\) and \(a^\dag\)). These commutation relations ensure that, in this locally flat coordinate frame, 
  \(a^\dag(k_3)\) and \(a(k_3)\) create and annihilate an amount of energy \(k^0(k_3)\) as usual. Thus,
particle states will be regarded as being produced from the vacuum by applying \(a^\dag\) to vacuum states, in the usual manner:\ \(H=k^0a^\dag a\)\,; \ \(k^0=\sqrt{k_3^{\,2}+\mu^2}\,\). 
 The general expression for the field \(\phi\) is such that \(A\) corresponds to annihilation and \(A^\dag\) to creation:
 \be \phi( Z,\,T)=A( Z,T)+A^\dag( Z,T)\ , \eel{PhiAA.eq}
 with %%
 \be A( Z,T)=\int\frac{\dd k_3}{\sqrt{4\pi k^0}} 
  \,a(k_3)\ex{i k_3 Z-ik^0\,T}; \labell {Afroma.eq} \quad \ \Ret
 \dot A( Z,T)=-i\int \dd k_3{\sqrt\frac{k^0}{4\pi }}
 a( k_3)
\ex{i k_3 Z-ik^0\,T}\ . 
 \eel{Adotfroma.eq}
  These equations show the creation and annihilation operators that can be used to generate any state with particles in the central region, as seen by a local observer. There will be upper and lower limits for the momenta \(k_3\).  The particles can be in region \(I\), where \(Z>0\) or in region \(II\), where \(Z<0\). So far, so good.

\newsecl{The Bogoljubov transformation}{bogoly.sec}
\subsection{Processing different annihilation and creation operators}
 Our point is now that in a locally flat space, the distinction between created and annihilated particles is fixed as soon as it is agreed upon what the time variable is. We need to replace the local time parameter \(T\) by the parameter \(\tau\) as it occurs in the metric  \eqn{rindler.eq}. This is a transformation on the time direction, which does not keep energy invariant. As we shall see, if one observer annihilates a particle, this may be regarded as a creation by an other observer.
 
 In the following, we make use  of the momentum parameters as used above:
 \be   k^\pm=\fract 1{\sqrt 2}(k_3\pm k^0)\ ,
 \quad k_3=\fract 1{\sqrt 2} (k^++k^-)\,, \quad k^0=\fract 1{\sqrt 2}(k^+-k^-)\,,\nm\\
  \hbox{obeying}\qquad  2 k^+k^-={k_3}^2-{k^0}^2=-\mu^2\ ,\quad \mu=  
 \sqrt{m^2+{\tl k}^2}\ .\eel{kkmu.eq} The light cone coordinates \(u^\pm\) are defined in Eqs~\eqn{loctrf.eq}.
 
 \subsection{Relating operator fields \(a\) to \(a_2\)}
 
At first sight, our problem seems to be easy. \(\tau\) is conjugate to \(\ln(u^+)\), or\(\ -\!\ln u^-\),
since an advance 
\(\del \tau\) in the \(\tau\) variable corresponds to multiplying \(u^\pm\) by \(e^{\,\pm\del\tau}\).
	Define\,\fn{Our notation follows the notation of Ref.~\cite{GtHSmatrix.ref} % 13
	, except that we  	skipped the use of the operator \(a_1\) there.}
\be a_2(\w)&=&\int_{0}^{\infty}\frac{\dd k^+}{k^+}\sqrt{\frac{k^0}{2\pi}}\,a(k_3)\,
		\ex{-i\w\ln \big(\fract {k^+\sqrt 2}{\mu}\big)}\ ,
\labell{a2wfroma.eq}\\ 
\hbox{with inverse:} \qquad  a(k_3)&=&\frac {1} {\sqrt{2\pi k_0}}
\int_{-\infty}^\infty\dd\w\,a_2(\w)\ex{\,i\w\ln \big(\fract {k^+\sqrt 2}{\mu}\big)} \ .\eel{awfroma2.eq}
The importance of this expression, the pre-factors, and in particular the factor \(\sqrt {k^0}\), can be understood as follows:

Note that, in Eqs.~\eqn{a2wfroma.eq} and \eqn{awfroma2.eq}, the integrals are one-dimensional, since we are keeping \(\tl k\) or \(\mu\) fixed (these do not participate in the transformations). Taking \(k^3\) as the only independent integration variable, and using Eqs.~\eqn{kkmu.eq}, we can write 
\be\dd k^+=\frac{\dd k^3}{\sqrt 2}\Big(1+\frac{k^3}{\sqrt{\mu^2+k_3^2}}\Big)=\frac {k^+\,\dd k^3}{k^0}
\ , \quad\hbox{so that}\quad \frac{\dd k^+}{k^+}=\frac{\dd k_3}{k^0}\ . \eel{k+k0.eq}
This allows us to derive from the definitions \eqn{a2wfroma.eq} and \eqn{awfroma2.eq}\,: \be [a_2(\w),a_2^\dag(\w')]=\del(\w-\w')\eel{a2comm.eq}
Of course,  the pre-factors in Eqs.~\eqn{a2wfroma.eq} and \eqn{awfroma2.eq} were chosen such that \eqn{a2comm.eq} is valid.

\subsection{Relating \(A\) to \(a_2\)}
 Now let \(A(Z,T)\) be  the part of the field \(\phi(Z,T)\) that is analytically proportional to the annihilation operators \(a\), see Eq.~\eqn{PhiAA.eq}. Then let us plug Eq.~\eqn{awfroma2.eq} into \eqn{Afroma.eq}, using \eqn{k+k0.eq}\,. Here is the detailed calculation\fn{Of course I do not expect that many readers will follow all these details; we wrote down the calculation to show that the Bogoljubov\cite{Bogoly.ref} %14
 transformation \eqn{eq28.eq} --- \eqn{eq32.eq} can be rigorously derived.}:  
\beq   &   A( Z,T)\ =\  \int\frac{\dd k_3}{\sqrt{4\pi k^0}}   \,a(k_3)\ex{+ik_3 Z-ik^0T} \   =\\&
=\ \int_{-\infty}^\infty\frac{\dd k_3}{2\pi k^0\sqrt 2} \int_{-\infty}^\infty{\dd\w}\ 
  		a_2(\w)\ex{\,i\w\,\ln\big(\fract {k^+\sqrt 2}{\mu}\big)+ik_3 Z-ik^0T}\   =   \\ &
=   \ \frac 1{2\pi\sqrt 2} \int_{-\infty}^\infty\dd\w   \int_{0}^\infty\frac{\dd k^+\,}{k^+}
	 	\,a_2(\w)\,\ex{\,i\w\ln\big(\fract {k^+\sqrt 2}{\mu}\big)+i \frac {Z}{\sqrt 2}\big({k^+}-
		 \frac{\mu^2}{2 k^+}\big)-i\frac{T}{\sqrt 2}\big({k^+}+
		 \frac{\mu^2}{2 k^+}\big)}\  =   \\ & 
=   \ \frac 1{2\pi\sqrt 2}
		 \int_{-\infty}^\infty\dd\w   \int_{0}^\infty\frac{\dd k^+}{k^+}
 		\,a_2(\w)\,\ex{\,i\w\ln\big(\frac {k^+\sqrt 2}{\mu}\big) +i\big({k^+}e^{-\tau}-
		  \frac{\mu^2}{2 k^+} e^\tau\big) \frac {\rr\vphantom{|_|}}{\sqrt 2}} \ = \\ &
=  \ \frac 1{2\pi\sqrt 2} \int_{-\infty}^\infty\dd\w   \int_{0}^\infty\frac{\dd k^+}{k^+}
		 \,a_2(\w)\,\ex{\,i\w\ln\big(\frac {k^+\sqrt 2}{\mu}\big)
		 +i\,\half \mu\rr -i\,\w\,\tau} \ .&  \eeql{atoa2.eq}				
Eqs.~\eqn{kkmu.eq} were used to write    
\be k_3=\frac {k^+}{\sqrt 2}\Big(k^+-\frac{\mu^2 }{2k^+}\Big) \,;\quad \hbox{and} \quad
k_0=\frac {k^+}{\sqrt 2}\Big(k^++\frac{\mu^2 }{2k^+}\Big) \,;\quad \ee 
Substituting  \(\fract{\sqrt 2}{\mu}\,k^+=s\,, \quad  
\alf=	\half \mu \rr\),\quad  we define a function \(F\):
 \be F(\w, \alf)&=&\int_0^\infty\frac{\dd s}{s}s^{\ds i\w}\,\ex{i\alf (s-1/s)}\,, \eel{Fdef.eq}
to rewrite Eq.~\eqn{atoa2.eq} as
\be A(Z,T)=\frac 1 {2\pi \sqrt 2} \int_{-\infty}^\infty\dd\w \,
a_2(\w)\,F(\w,\half  \mu \rr)\ex{-i\w \tau}\ .\eel{AZtau.eq}
We see clearly that, in terms of the annihilation operator \(a_2(\w)\), the \(\tau\) parameter acts as time, and \(\w\) represents the energies being annihilated. 

\subsection{Properties of the function \(F\)}
Properties of function \(F\) are easy to derive:
\be F^*(\w,\alf)&=&F(-\w,-\alf) \,, \labell{eq23.eq}\\
F(-\w,\alf)&=&e^{-\pi\w}F^*(\w,\alf)\qquad\hbox{if}\qquad \alf>0,\ \labell{eq24.eq}\\
F(-\w,\alf)&=&e^{+\pi\w}F^*(\w,\alf)\qquad\hbox{if}\qquad \alf<0\, .\eel{eq25.eq}
The last two equations follow from replacing \(s\leftrightarrow  s\,\ex{\lam i}\). In Eq.~\eqn{eq24.eq},
the rotation \(\lam=0\) to \(\lam =\pi\) makes the integral convergent only if \(\alf>0\) \,. To prove Eq.~\eqn{eq25.eq}, we rotate the \(s\) variable in the opposite direction.

Eq.~\eqn{AZtau.eq} is crucial for understanding the Hawking radiation phenomenon. The \(\w\tau\) combination 
found there implies that the operator \(a_2(\w)\) annihilates an amount \(\w\) from the energy in the field \(A\).
 This is the energy of the light cone variables in region \(I\) of the Penrose diagram,  see Figure \ref{SS-KS.fig}.   

\subsection{How \(a_2\) and \(a_2^\dag\) mix to form \(a_I\) and \(a_{II}\)}
However, now comes the crucial point. In Eq.~\eqn{awfroma2.eq}, the integration goes from \(\w=-\infty\) to \(+\infty\), so that \(\w\) can be negative. An operator that annihilates a negative amount of energy, is actually a creation operator, and should be handled as such. We see that the integrals in Eq.~\eqn{atoa2.eq} have to be split in two parts, depending on the sign of \(\w\).

In short: in region \(I\), \(a_2\) is an annihilation operator if \(\w>0\) and a creation operator if \(\w<0\).
In region \(II\) it is the other way around. However,  also another step is needed:

An observer in region \(I\) can only observe the fields in region \(I\). These are the fields where the coordinate \(\rr>0\). Thus, in region \(I\) we have the annihilation operators that have \(k^0>0 \) and \(\w>0\). Etc. Therefore, we split the integrals \eqn{a2wfroma.eq} and \eqn{awfroma2.eq} in accordance with the sign of \(\w\), and the sign of \(\rr\). Consider the fields in region \(I\) and take \(\rr>0\)\,. Using the properties \eqn{eq23.eq}--\eqn{eq25.eq}, we rewrite \eqn{AZtau.eq} as
\be &\phi(\rr,\,\tau)=A(\rr,\,\tau)+A^\dag(\rr,\,\tau)=\labell{eq26.eq}\\
&=\ds \int\frac{\dd k_3}{2\pi\sqrt{2}}\int_0^\infty\dd\w F(\w,\half\mu \rr)\Big(
a_2(\w)+e^{-\pi\w} a_2^\dag(-\w)\Big)+h.c.&\eel{eq27.eq}  %%%%
Here, the terms inside the large brackets are to be associated with \(A(\rr,\,\tau)\) and their hermitian conjugates to \(A^\dag(\rr,\,\tau)\). This assures that all terms for \(A(\rr,\tau)\) annihilate positive amounts of energy, while \(A^\dag(\rr,\tau)\) creates positive energy. 

Note that in Eqs~\eqn{eq26.eq} and \eqn{eq27.eq}, \(\rr>0\).  The expressions inside the large brackets always enter in this combination, as long as we are dealing exclusively with the operators defined in region \(I\). 

We see that, if we restrict ourselves to the fields in region \(I\) only, we find that they do generate pure annihilation operators, but only in this  combination: the operator \(a_2^\dag(-\w)\)  is to be added with the reduction factor \(e^{-\pi\w}\), which originates in Eqs.~\eqn{eq23.eq}--\eqn{eq25.eq}.
This mixture of creation and annihilation operators is characteristic for a \emph{Bogoljubov transformation}\cite{Bogoly.ref}.
Take \(\w>0\)\,. We define \(a_I(\w)\) in region \(I\) and and \(a_{II}(\w) \) in region \(II\):
\be a_I(\w)=\NN\big(a_2(\w)+e^{-\pi\,\w}a_2^\dag(-\w)\big)\,.\labell{eq28.eq}\\
a_{II}(\w)=\NN\big(a_2(-\w)+e^{-\pi\,\w} a_2^\dag(\w)  \big)\, .\eel{eq29.eq}
Here, \(\NN=1/\sqrt{1-e^{-2\pi\w}}\) is a normalisation factor, to ensure that 
\be [a_I(\w),\,a_I^\dag(\w')]=[a_{II}(\w),\,a_{II}^\dag(\w')]=\del(\w-\w')\,.\eel{eq30.eq}

The inverse of this set of equations are:
% \,\fn{The signs in these equations may seem to be confusing, unless one realises that 
%  these are just very ordinary two-dimensional hyperbolic orthogonal rotations:
 % \(\NN=\cosh\vv\,;\ e^{-\pi\w}=\tanh\vv\,.\)} 
(again when \(\w>0\))\,:
\be a_2(\w)=\NN\big(a_I(\w)-e^{-\pi\,\w}a_{II}^\dag (\w)\big)\,,\labell{eq31.eq}   \\
a_2(-\w)=\NN\big(a_{II}(\w)-e^{-\pi\,\w}a_{I}^\dag(\w)  \big)\,. \eel{eq32.eq}

For a local observer to whom the region at the origin of the Penrose diagram seems flat, all operators \(a_2(\w)\) are superpositions of the annihilation operator \(a(k_3)\)\,, for all \(\w\), see Eqs.~\eqn{atoa2.eq}.
On the other hand, for an observer who only has access to region \(I\), only the 
operators depending on \(\phi(\rr,\tau)\) with \(\rr>0\) are visible.  These are the operator combinations \(a_I(\w)\) with \(\w>0\)\,.

What are the physical consequences to be deduced from these transformation rules?
\newsecl{Hawking's original result in this formalism}{Hawk.sec}
\subsection{The vacuum state \(|\W\ket\) for the local observer}
	All states with or without particles in them, can be obtained by allowing creation operators, such as 
	\(a^\dag(k_3)\),  \(a_2^\dag(\w)\), \(A_I^\dag(\rr,\tau)\) or \(a_{II}^\dag\) to act on a vacuum state. Let us 	
	start from the vacuum state \(|\W\ket\) as experienced by an observer near the center of the Penrose 
	diagram. For this observer, the annihilation operators \(a(k_3)\) and \(a_2(\pm\w)\) all take the value zero, 
	as there is nothing to annihilate:
\be a(k_3)|\W\ket =0\, ;\quad\hbox{therefore} \quad  a_2(\pm \w)|\W\ket =0\ . \eel{aOmegaket.eq}	
	This enables us to use Eqs.~\eqn{eq31.eq} and \eqn{eq32.eq} to derive all elements \(\bra n_I,\,n_{II}|	\W\ket \) up to an overall constant:
	
we find that 
\be a_I (\w)|\W\ket = \ex{-\pi\w} a_{II}^\dag|\W\ket\ , \quad\hbox{and}\quad a_{II}(\w)|\W\ket=\ex{-\pi\w}a_I^\dag(\w)|\W\ket\ . 
\eel{eq35.eq}
Since all annihilation operators \(a(\w)\) act  on a state with \(n\) \(\w\)-objects as 
\be a(\w)|n\ket=\sqrt{n}\,|n-1\ket\,,\quad\hbox{or}\quad \bra n-1|a(\w)|\W\ket = 
\sqrt n \bra n|\W\ket \ ,\eel{eq36.eq}
we derive for all basis elements where \(n_I=n_{II}=n\ge 0\)\ :
\be \bra n+1,\,n+1\,|\W\ket = \ex{-\pi\w}\bra n,\,n\,|\W\ket \ . \eel{eq37.eq}
In this case, the factors \(\sqrt {n(\w)}\) cancel out. They do not cancel out if we apply Eqs.~\eqn{eq35.eq} to states where \(n_{II}\ne n_I\)\,. In that case, Eq.~\eqn{eq28.eq}
gives
 \be \bra n+1,m|\W\ket =\sqrt{\fract m{n+1}}\,\bra n,m-1|\W\ket\ex{-\pi\w} \,; \eel{eq42.eq}
using this recursively would lead to states with negative \(n_I\) or \(n_{II}\)\ , but for \(n<0\) we cannot use Eq.~\eqn{eq42.eq}. Thus, we arrive at
\be \bra n_I,\,n_{II}|\W\ket = \bra 0,\,0|\W\ket \ex{-n_I\,\pi\w}\ \del_{n_I,\,n_{II}}\ , \eel{eq38.eq}
where \(\bra 0,\,0|\W\ket=\sqrt{1-\ex{-2\pi\w}}\)\,, since \(|\W\ket\) must be normalised: \(\bra\W|\W\ket =1.\)

The Hamiltonian that measures the energy associated to advances in the time-coordinate \(\tau\), can be written as
(note the sign of \(\w\)):
\be H_{\tau}=\int_{-\infty}^\infty\dd\w\,\w a_2^\dag(\w)a_{2}(\w)-C(\w)=
\int_0^\infty\w\dd\w\Big(a_I ^\dag(\w) a_{I}(\w)-a^\dag_{II}(\w)a_{II}(\w)
\Big)\,.\eel{eq40.eq} 
%temp
This identity follows directly from Eqs.~\eqn{eq31.eq} and \eqn{eq32.eq}.  \(C(\w)\)
 is a constant that tunes the energy of the vacuum to zero, through normal ordering.  From here on, 
 take \(\w>0\)\,, and we define the signs of the energy operators \(H_I\) and \(H_{II}\) both to be positive.
 
 We then see that region \(II\) contributes negatively to what seems to be 
the total energy. This happens because the time evolution as a function of \(\tau\), in region \(II\), goes in the opposite direction, see Fig.~\ref{KSCauchy.fig}b. 
 
 \subsection{The generator of local Lorentz transformations\labell{lolo.sub}}
 The Hamiltonian for the distant observer can be seen to be the generator of Lorentz transformations in the radial direction for the local observer. This means that, if \(L\) is the local Lorentz generator, \(H_I\) is the Hamiltonian for the distant observer in region \(I\), and \(H_{II}\) is the (positive) Hamiltonian in region \(II\), whereas
  \be L=H_\tau=H_I-H_{II}\ . \eel{Lorentz.eq}
 
The operators \(a_2(\pm\w)\) give zero because these are linear functions of \(a(k_3)\), see Eqs.~\eqn{Afroma.eq} and \eqn{awfroma2.eq}. But the operators \(a_I(\w)\) and \(a_{II}(\w)\) do not vanish, since they receive admixtures of \(a^\dag(k_3)\) (or equivalently, \(a_2^\dag(\w)\)), see Eqs \eqn{eq28.eq}--\eqn{eq32.eq}).  The operator  \(a_I(\w)\) only annihilates particles in region \(I\), and \(a_{II}(\w)\) annihilates only particles in region \(II\). 

The operator \(L\) is the local generator of the Lorentz transformations and as such more basic 
than the operator \(H_I+H_{II}\).

Now consider a given value for \(\w\).  If \(a_I(\w)\) and \(a_{II}(\w)\) do not vanish, this must mean that the observer in regions \(I\) or \(II\) do observe particles. let us calculate what they see. Given \(\w\), write  a state containing \(n_1\) particles in region \(I\) and \(n_2\) particles in \(II\), as 
\(|n_1,\,n_2\ket\), \ or \ \(|n_1\ket_I\  |n_2\ket_{II}\).

\(|\W\ket\) is defined by~\eqn{aOmegaket.eq} and \eqn{eq38.eq}, which is Lorentz invariant. Therefore, one may expect \(|\W\ket\) itself to be Lorentz invariant. Consequently, for all \(\w\), we must have: 
\be &&L|\W\ket=(H_I-H_{II})|\W\ket=0\ ;\labell{Lzero.eq}\\
&&\bra n_1,\ n_2|L|\W\ket=(n_1-n_2)\bra n_1, n_2|\W\ket=0\ ;\eel{n1n2Omega.eq}
hence,  \,\(\bra n_1,n_2|\W\ket=0\) \, unless \(n_1=n_2\). This is what we derived in Eq.~\eqn{eq38.eq}. We see here that it is to be interpreted by observing that the state \(|\W\ket\) must be Lorentz invariant. 

\subsection{The emergence of Hawking radiation, Scenario 1}
This derivation leads to an important physical result: \emph{A distant observer experiences the apparent emergence of particles, both in region \(I\) and in region \(II\). } These particles are neatly described by a wave function that puts equal numbers of particles in region \(I\) and region \(II\). The state \(|\W\ket\), which can also be considered for an entirely flat space-time, is called the Unruh state.\cite{Unruh.ref}. %15

Adding a bit of physical intuition, one now continues to ask the question:  suppose we have an observer in region \(I\), who has no access to region \(II\). According to the local observer, the state \(|\W\ket\) is the only time-translation invariant one. What would the observer at infinity see? The answer was given immediately. In nearly all treatises about quantum black holes, we find an argumentation, that, in our notation,  
amounts to \emph{scenario 1}:

The wave function for \(n_1\) particles in region \(I\) and \(n_2\) particles in region \(II\) is 
		\be\psi(n_1,\,n_2)=C^{st}\,\ex{-n_1\w\pi}\del_{n_1,n_2}\ .\eel{psin1n2.eq}
This is a quantum state, which needs to be normalised to one\fn{We repeated the calculation of expression \eqn{eq38.eq} here, because it is essential to understand where it comes from.}:
	\be \sum_{n_1,n_2}|C^{st}|^2\,\ex{-2\pi n_1\w}\del_{n_1,n_2}=|C^{st}|^2\sum_{n=0}^\infty\ex{-2\pi\w n}=1\ .\ee
	Now, since the sum here is \((1-\ex{-2\pi\w})^{-1}\), we find
\be  |C^{st}|^2=1-\ex{-2\pi\w}\ . \ee	
In region \(I\) one cannot observe the particles in region \(II\). Therefore, the probability to observe \(n_I\) particles in region \(I\) is the absolute square of this wave function, \emph{summed over all unobserved states \(|n_{II}\ket\) in region \(II\)\,.}   But in the state \(|\W\ket\), the number \(n_2\) can take only one value: \(n_2=n_1\). Therefore,  the probability  that, given the wave function \(|\W\ket\), we see \(n\) particles in region \(I\) is
	\be|C^{st}|^2 e^{-2\pi n\w}=(1-e^{-2\pi\w})\,e^{-2\pi n\w}\eel{stationarystate.eq}
This is a typical thermodynamical expression, and, since the energy of an \(n\) particle state is \(n\w\),  the temperature is given by
\be \bet=2\pi\,,\quad\hbox{or, we have a temperature}\quad T=1/(2\pi)\,,\eel{Htemp.eq}
in the units that fit with the given black hole mass. Replacing the time unit \(\tau\) by \(t/4GM\), Eq.~\eqn{rindler.eq}, we get the Hawking temperature,
\be T_H\qu 1/8\pi GM\ . \eel{Tqu.eq}

A peculiar phenomenon is that, if one \emph{would} have access to both regions, one would find that the particles in region \(I\) and region \(  II \) are precisely correlated: if, with the appropriate thermal suppression factor, a heavy particle  or state would have been observed in \(I\), then, for certain, a heavy particle would be guaranteed to emerge also in region \(II\), while the expected additional thermal suppression factor there, would be totally absent.

And yet, we put a question mark in Eq.~\eqn{Tqu.eq}. Not everything is right. We calculated that the number of particles emerging in region \(II\) always equals the number emerging in region \(I\).  How do the particles in universe \(II\) know what the number \(n\) is in region \(I\)?

\newsecl{Thermal density matrix and radiation temperature}{temper.sec}
\subsection{Scenario 2}
The calculation of the state of the particles emerging from the black hole, Eq.~\eqn{eq38.eq}, seems fine, technically. But it is the quantum state \eqn{psin1n2.eq} that we find here, in `scenario 1' , that leads to  concern\fn{The observation that different approaches yield different Hawking temperatures was first made in Ref.\,\cite{GtHambi.ref} %16
.}. 

The particles in region \(II\)  obey field equations exactly identical to the ones in region \(I\). Their numbers are always equal to the numbers in region \(I\): \ \(n_{II}=n_I\) . Why are they in a `hidden' universe?

This author claims that they are not hidden. A perfect scenario can be arranged, which puts everything in a frame that should look familiar. This we call scenario~2: \emph{thermodynamics in a single universe}. Look at the evolution equations. The Hamiltonian \(H_{II}\) controlling region \(II\) is mathematically perfectly equal to the Hamiltonian \(H_I\) controlling region \(I\), except for one thing: according to the laws of the time evolution, where the variable \(\tau\) acts as time, the Hamiltonian \(H_{II}\) should be \emph{minus} \(H_I\). We already saw, in Eq.~\eqn{Lzero.eq}, that Lorentz invariance connects \(H_I\) with \(-H_{II}\). A time translation in \(\tau\) sends all wave functions in region \(II\) to the minus direction. In short, the wave function for the entire system evolves with the total Hamiltonian \(H_I-H_{II}\). 

But this is nothing new. This is how the density matrix, \(\rr(\tau)=|\psi_\tau\ket\,\bra\psi_\tau|\) evolves. \(|\psi_\tau\ket =\ex{-iH\tau}|\psi_0\ket\) is the ket state, and \(\bra \psi_\tau|=\bra \psi_0|\ex{iH\tau}\) is the bra state. This seems to make sense. 
Indeed, Hawking\cite{Hawkdens.ref}  has been exploring exactly this alley in his earlier investigations concerning the information loss problem. In this scenario, nothing about the wave function gets lost; all we have to do is figure out how the black hole evolution looks from this vantage point. % Hawking density matrix?

Indeed, there is a consequence of this that was pointed out by this author long ago\,\cite{GtHambi.ref};  it appears to have enjoyed little attention. If  the density matrix \(\rr\) is  directly described by the product of states in region \(I\) and states in region \(II\), we have, in the stationary case,
\be |\W\ket= \rr=\sum C^{st}|\psi\ket_{I\ II}\bra\psi| \,, \eel{rho=psi2.eq}
where \(C^{st}\) is a normalisation factor.
Indeed, the state \(|\W\ket\), Eq.~\eqn{psin1n2.eq} has exactly this form:
\be |\W\ket=C^{st}\sum_{n\ge 0}|n\ket \ex{-n\pi\w}\bra n|\ .\eel{eq72.eq}
However, if we now calculate the temperature, using the Boltzmann-Maxwell distribution, we get
\be\Tr\big(C^{st}\sum_n|n\ket \ex{-n\pi\w}\bra n|\big)=C^{st}\sum_n\bra n|\ex{-n\pi\w}|n\ket=\sum_n\bra n|\ex{-\bet n\w}|n\ket\ . \eel{eq73.eq}
It follows that the inverse temperature \(\bet\) in this scenario, scenario 2, is exactly half the value  \eqn{Htemp.eq} obtained in scenario 1. 
If now we derive the Hawking temperature, we get 
\be T=1/4\pi G M\ ,\eel{2T.eq}
This must look confusing to the reader. Expression \eqn{eq72.eq} looks exactly like Eq.~\eqn{psin1n2.eq}. Whence this factor 2 in the temperature?

\newsecl{Region II as quantum clone of region I}{regions.sec}
\subsection{The difference between the two scenarios}
We started with the same expressions for the Bogoljubov-transformed vacuum state \(|\W\ket\) both in Section \ref{Hawk.sec} (scenario 1), and in \ref{temper.sec} (scenario 2). Yet the temperature we computed differs by a factor 2. Whence this factor? One answer is that the wave function had to be squared in scenario 1, while we used it linearly in scenario 2. This lead to the term \(\ex{-2\pi n\w}\) in scenario 1, while we had \(\ex{-\pi n\w}\) in scenario 2, and showing this was the reason for doing the calculation as accurately as possible in sections \ref{Hawk.sec} and \ref{temper.sec}.

But where did this difference arise?  And which of the two calculations, if any, is correct?

The difference comes about in our use of the inner products. In Section~\ref{Hawk.sec}, we handled the states \(|n\ket\)  in region \(II\), ( or \(\bra n|\)) as all being orthogonal to all states \(|n\ket\). In Section \ref{temper.sec}, we use the inner products \(\bra n|n'\,\ket\) as non vanishing if \(n'=n\). Of course: the inner product \(\bra n|n'\ket=\del_{n,n'}\) when used in the density matrix.

The conclusion we do arrive at is that region \(II\) is filled with particles exactly the same way as region \(I\), apart from the fact that region \(II\) evolves not exactly the same way as region \(I\), but as the \emph{hermitian conjugate} of  \(I\). 

Since probabilities are now obtained directly from a density matrix, rather than a wave function, one might even suspect that this result agrees well with a probabilistic interpretation of quantum mechanics in general. This also does not prove who is right, but we have not yet used all our ammunition. There is something else to worry about: does the quantum evolution preserve information? Do our methods, as used so-far, give us   clues for other approaches?

\begin{figure}\widthfig{220pt}{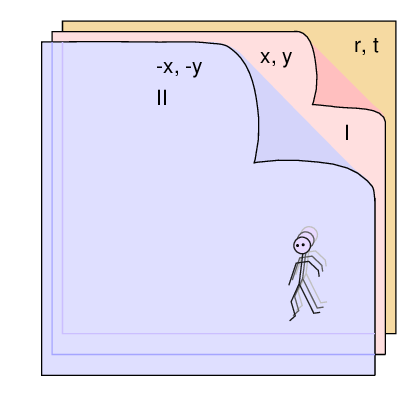}\caption{\small Region \(II\) as an exact quantum clone of region \(I\). Not only all particles are the same ones in the two regions, also the observer is copied. In this scenario, it is better to return eventually to the Schwarzschild metric (\(r,\,t\)), of which both region \(I\)  (\(x,\,y\)) and region \(II\) (\(-x,\,-y\)) are obviously quantum copies of just one world, specified by the unique coordinates \(r\) and \(t\).\labell{clones.fig}}
\end{figure}

\subsection{Rindler Space}
Rindler space\,\cite{Rindler.ref} is a description of ordinary flat space time using a parameter space where time is replaced by the generator \(\tau\) for Lorentz transformations in the \(z\)-direction. Here, the physics is well understood, and we emphasise that scenario 1 applies to Rindler space: the temperature detected by an observer who is accelerated with respect to the horizons of this space-time, has always been calculated correctly. 

Rindler space hides  two entire universes behind its two horizons, and, in this case we understand precisely how to do the calculation: all ket states \(|\psi\ket\) in region \(I\) are to be considered orthogonal to all bra states \(\bra\psi|\) in region \(II\), so that, in particular, the inner products \(\bra\psi| \psi\ket\) vanish, unlike the way they are used in Eq.~\eqn{eq73.eq}. The two universes represented in Rindler space may be entirely different, so
they will in general not be identical twins. One can have a solar system in one of them and a totally different alien world in the other. The observer has to decide in which of the two universes to sit when performing measurements. 

The two regions exhibited in scenario 2 are identical twins --  they are equal in the quantum mechanical sense:  quantum clones of one another. This situation is illustrated in Figure~\ref{clones.fig}. Every visible space-time point in the Schwarzschild metric, marked \(r,\,t\), corresponds to a point in region \(I\), marked \(x,\,y\), and it also corresponds to a point in region \(II\) marked \(-x,\,-y\). Even the observer is represented in these three frames. This is why we call these three frames quantum clones.

We end this section by anticipating how we will be able to decide what model for region \(II\) to use for describing a physical black hole. The extended description of the black hole metric, the Kruskal-Szekeres coordinate system (section \ref{KS.sec}, fig.~\ref{SS-KS.fig}b and fig.~\ref{KSCauchy.fig}), shows two regions, and as yet, we did not have any use for the particles in region \(II\) of this metric. We claim that black holes must be fundamentally different from Rindler space; Rindler space can only be described by scenario 1, while black holes can only be described by scenario 2, if we may assume that they preserve quantum information.

\subsection{Clones}

What we have so-far is that, if there are \(n\) particles in region \(I\), there will also be \(n\) particles in region \(II\). This comes in addition to the observation that, if we consider the two regions \(I\) and \(II\) in \emph{any} black hole solution without matter, we always find these two regions to be identical.  Actually, we are only able to admit exactly as many matter particles in region \(II\) as we have in region \(I\). This could be regarded as a law of Nature: \emph{any modification} in region \(I\) of a black hole, is to be accompanied by an \emph{identical} modification in region \(II\).

Or: there may be \emph{no way} to bring about even the slightest difference between region \(I\) and region \(II\). As we already stated, it implies that region \(I\) and region \(II\) are \emph{quantum clones} of one another. This means that \emph{even the outside observer} is copied in region \(II\) as soon as he or she pops up in region \(I\);  see Fig.~\ref{clones.fig}. A striking consequence of this law will be that region \(II\) will contain no information whatsoever that is not already available in region \(I\). Actually, there is a simpler way to phrase this law: Regions \(I\) and \(II\) also carry a unique copy in the original Schwarzschild coordinate frame. Since the Schwarzschild coordinates do not allow different copies of themselves, there are no other copies.  Note that we do not have such a law in Rindler space.

The importance of this carbon-copy law is now two-fold: one, it may well enable  us to understand how information can be preserved in an evolving black hole (information has no way to hide since all information in region \(II\) is already visible in region \(I\), and in the Schwarzschild frame. This may well enable us to dismiss one or both of the scenarios sketched in sections \ref{Hawk.sec} and \ref{temper.sec}. 

Second, since all contents of region \(II\) are now obviously observable in region \(I\), one may already argue that information sent into a black hole has nowhere to hide. We shall see how this point materialises in the next sections.

\subsection{A conical singularity} Perhaps the situation can be clarified a bit more if we use a different language.  Identifying one region of a metrical space with another region, is tantamount to dividing this space-time by \(\ZZ(2)\), and the physical effect this has can be described in terms of a conical singularity. Here, the conical singularity is placed at the center of the 
Kruskal-Szekeres metric. Since the time coordinate is involved in an essential manner, one can easily imagine how this modification of the metric also generates a factor 2 in the temperature. Careful inspection indicates that the unit of time used far from the center differs by a factor 2 from the unit used initially, and, as temperature has the dimension of energy, or the inverse of the time unit,  the temperature is modified by a factor 2.

Conical singularities are physically observable, so what we have here is a modification of the local laws of physics! it may be worth-while to investigate how the existence of such singularities may affect the Standard Model. This has not yet been done, as far as we know.

\newsecl{The Shapiro effect linking in- and out-particles}{shapiro.sec}
\subsection{How in-particles affect out-particles and vice versa, on the horizon}
One of the main topics in this book, is the question how matter going in (`in-particles'), may be expected to transfer the information they carry, onto particles going out (`out-particles'). Now as long as these particles stay away from any of the horizons, we may assume that a field theory can be used such as the Standard Model, modified so that it can be applied on a modestly curved background metric.

At first sight, one might expect that the effects that in-particles have on out-particles, is far too weak to transmit \emph{all} information from one to the other. Certainly a graviton hopping over should be ineffective.

But then one underestimates this force; note that gravitation scales, to become strong at tiny distances. More to the point, the effect has been calculated. It is found that the curvature of the background metric will cease to be modest when visited by particles \emph{on} one of the horizons. Unfamiliar, but computable,  phenomena take place when an in- particle crossing a horizon also crosses  the orbit of a particle moving out in a geodesic parallel to that horizon. What happens can be entirely calculated using standard, non-quantum mechanical equations: Einstein's field equations. All we need are the momenta and the positions of these particles that meet one another. The calculation is completely clean, and can be done with relative ease. The way to discover this  is displayed in detail in Ref.~\cite{GtHSmatrix.ref}. Here, we briefly summarise the main features to enable the reader to follow the developments on this topic.

\begin{figure}\qqqquad 
a)\widthfig {106pt}{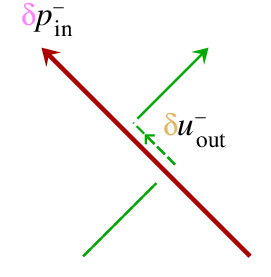}\qqquad b)\widthfig{120pt}{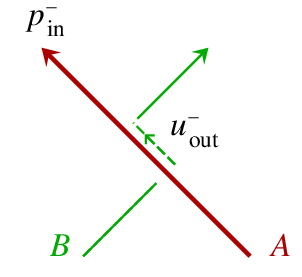}
\caption {\small A particle with high momentum (moving to the left) crosses the orbit of a low momentum particle.\labell{AS.fig}
\ a) The position of the right moving particle is changed. \ b) By using appropriate coordinates, the shift may be regarded as an absolute coordinate.}
\end{figure}

We use the feature described in Ref.~\cite{ASx.ref}--\cite{GtHDray.ref} %18
. It is illustrated in Fig.~\ref{AS.fig}.
With almost the speed of light,  a high-energy particle \(A\) is crossing the orbit of a light test particle, \(B\).
The particle \(B\) then experiences a time lapse. The magnitude of the time lapse 
was calculated by Shapiro\cite{Shapiro.ref} %19
, deriving how radio and optical radiation is slowed down when signals are grazing past the Sun. The Sun does to the radio signals  what particle \(A\) does to particle \(B\). Since the gravitational field is controlled by field equations, the effect on the spectator particle \(B\) is described by a Green function.

\subsection{The Green function on the horizon}

Let particle \(A\) (later to be chosen to be the in-particle) have 4-momentum \(p_A\), which dominates in the light cone direction. 
The gravitational field of particle \(A\) is here Lorentz contracted so strongly that the effect is 
limited to a light-like plane, and the spectator particle \(B\) is dragged along by a distance \(u_B\) in the 4-direction of the motion \(p_A\) of \(A\).
The magnitude of the drag is given by
\be u_B(\tl x)=\int\dd^2\tl y\,f(\tl x-\tl y)p_A(\tl x)\ , \ee
The Green function \(f\) lives in the two dimensional plane orthogonal to \(p_A\) and \(p_B\).

If we Lorentz transform the entire scenery such a way that particle \(A\) will almost stand still, we find that not \(B\) 
but \(A\) is being dragged along by exactly the same mechanism. The phenomenon was described by Aichelburg and Sexl\cite{ASx.ref}, and Bonner \cite{Bonnor.ref}%20
.
Applying this to particles near a black hole horizon we see the same phenomenon, where now the horizon plays the same role as   the flat plane in the set-up by Aichelburg and Sexl, but now this space is a 2-sphere. On this 2-sphere, the Green function \(f\) takes a form given by the equation
\be (1-\Del_\W)f(\W,\W')=8\pi G\del^2(\W,\W')\ , \eel{eq92.eq}
so that 
\be u^\outt(\W)=\int\dd^2\W'\,f(\W,\W')p^\inn(\W'),\eel{eq93.eq}
and conversely,
\be u^\inn(\W)=-\int\dd^2\W'\,f(\W,\W')p^\outt(\W')\ . \eel{eq94.eq}
We used a notation where the momentum of the in-particle is denoted as a distribution, 
usually of the form\, \(p^\inn(\W)=p_1\del^2(\W,\,\W_1)\).
The derivation of equation \eqn{eq92.eq} is actually  a rather tedious one. It was derived in
% Kijken of we dit nog kunnen verduidelijken.
Ref.~\cite{GtHSmatrix.ref}.
\subsection{Information retrieval}
Using this phenomenon, we shall describe the process of information retrieval for black holes 
in the remainder of this chapter.  The in-going matter is assumed to be described by a distribution of momenta, generated by many particles going in, but not so many that we need a second quantised procedure\fn{One expects in the order of one Hawking particle to be emitted in a period of time in the order of \(1/ G M_{BH}\).}.   Consider \(N\) particles going in, enumerated by \(k=1, \cdots,\,N\). The in-momentum distribution is then given by the sum
\be p^\inn(\W)=\sum_{k=1}^N \del^2(\W-\W_k)p_k\ . \ee
It must be noted that this is not exactly the form in which the Standard Model particles arrive at the horizon; in this expression, they may be viewed as second-quantised particles only w.r.t. the 2-sphere of the horizon, while,  as a function of the radial coordinate \(\rr=r-2GM\), we only see the single operator \(p^\inn\). 

We anticipate that this mismatch can be handled, but, we emphasise that the presentation that we advertise now, will be unitary, in a way to be explained, which is why we here briefly summarise how this is done. 

The momentum distributions \(p^\inn(\W)\) and the coordinate shift \(u^\outt(\W)\) are both expanded in spherical harmonics:
\be p^\inn(\W)&=&\sum_{\ell,m}p_{\ell,m}^\inn Y_{\ell,m}(\W)\ ;\labell{eq99.eq}\\ 
u^\outt(\W)&=&\sum_{ell,m}u_{\ell,m}^\outt Y_{\ell,m}(\W) \ . \eel{eq97.eq}
The point is that these particles do not represent the entire set of all particles ever absorbed by the black hole, but only those particles that are seen in a given time interval, representing a stretch of time roughly of the order of \(GM\) in natural units, so they are particles that describe only small deviations 
in a Schwarzschild metric that, by itself, is perfectly time translation and rotation invariant. 

We claim that the disturbances caused by much earlier in-going and much later out-going particles will be represented by the particles that are to be described in the other time sectors, and the equations we derive will show how this happens.\fn{The Page time\,\cite{Page.ref}, often introduced in this context, is a far too long time stretch to be of use here, as explained in a footnote in Section~\ref{globallocal.sec} of this chapter.}

Now consider Eqs.~\eqn{eq92.eq} and \eqn{eq93.eq}. In the spherical wave expansion, these take the form
\be {u^-}^{\,\outt}_{\ell m}=\frac{8\pi G}{\ell^2+\ell+1}p_{\ell m}^{-\,\inn}\ ,\eel{eq98.eq}
where the superscript \(-\) stands for the \(-\) direction in the local Lorentz frame close to the horizon.

The Green function  diagonalises when using the partial wave expansion, and hence 
\({u^-}^\outt_{\ell m}\) has become detached from the other partial wave components 
\((\ell\,', m')\), and, consequently, our system is simplified enormously by this procedure\,\fn{Our approach was criticised by claiming that transverse components of the gravitational fields should invalidate the approximations. But, while the transverse components of the momenta are basically time-independent, the radial components increase exponentially with time. This is why we consider these as being dominant. Only at large \(\ell\) values, we may expect stronger transverse perturbations, for which a cut-off must be assumed. We expect that large \(\ell\) values will not dominate, but the side effects at large \(\ell\) still have to be sorted out.} This is in fact the same procedure as the one that enables the calculations of the electron wave functions for the hydrogen atom. The classical photon field in the atom is the analogue of the classical graviton field in our equations here.

But there is something more. Eq.~\eqn{eq98.eq} is a simple operator expression that can be seen to act on quantum states \(|{p^{-\,\inn}_{\ell m}}\ket\) and \(|u^{-\,\outt}_{\ell m}\ket\), which we may subject to Fourier transformations to get 
\( |u^{+\, \inn}_{\ell m}\ket \) and \(|{p^{+\outt}_{\ell m}}\ket\ . \) This gives exactly  Eq.~\eqn{eq94.eq}, with the minus sign.

Indeed,  there is a neat way to employ quantum commutation rules to see that the information carried by the in-particles takes the form of a Hilbert space that returns exactly the same information to the out-particles.  

Note however, that this statement only holds for the two regions, \(I\) and \(II\), combined. How to separate the information for the two different regions, is an important question, to be meticulously  addressed in the next sections.		% will it?

Consider the shifts in the coordinates \(u^-\) caused by the in-particles with momenta \(p_k^\inn\). These momenta increase exponentially in time, which is why the shifts \eqn{eq98.eq} they cause are essential.  
The total of the in-going momentum is a distribution \(p^-(\W)\) where the effects of all in-particles are added together. In turn, they shift all out-particles, at different values of \(\W\), by the amount \(u^-(\W)\). Since the effects of early in-particles on late out-particles diverge exponentially with time difference, we only take into account the particles inside the time zone indicated by the green region in Fig.~\ref{SS-KS.fig}a. How to handle the other particles will also be addressed. % will it?

	We have to assume that all in-particles and all out-particles arrive or leave at different values for the angles \(\W\) (which, in fact, is a good approximation; for large black holes, the odds against two particles entering or leaving at exactly the same values of \(\ell\) and \(m\) is negligible). Then we see that Eq.~\eqn{eq98.eq} in fact suffices to relate all out-particle information to the information from the in-particles. The relation is in the form of a Fourier transformation. Fourier transformations preserve the norm of a function, and therefore, they are unitary.

\newsecl{More about the clones}{clones.sec} 
	%\qqquad titel lijkt te veel op die van sect 8 !!!\\
	\subsection{Out-particles are the Fourier transforms of the in particles}
	In this section we shall derive further support for the factor 2 that was first encountered in Section~\ref{temper.sec}, the fact that the temperature of Hawking radiation must be given by Eq.~\eqn{2T.eq} rather than \eqn{Tqu.eq}. One thing that one might conclude from the previous section is, that we did not yet have everything completely right. 

Only the \emph{geometric} properties of the in-particles are neatly transmitted by the \emph{geometric} properties of the out-particles, but how non-geometric properties are to be handled is not yet understood. We leave this here for future investigations. 

However, if, by way of simplification, we limit ourselves to scalar, massless particles entering and leaving the black hole, much more can be understood.

To do this analysis right for the scalar particles,  we first need to achieve optimal  understanding of the geometric parts (momenta \(p^\pm_{\ell,m}\) and positions 	\(u^\pm_{\ell,m}\)). By investigating the gravitational effects of in-particles onto out-particles, we find that the out-\emph{positions} are determined by the in-\emph{momenta}, and \emph{vice versa}. Quantum mechanics clearly states that momenta are the Fourier transforms of the positions, and \emph{vice versa}. The Fourier transformation operator is unitary since the norm of a function is preserved. This is the first step towards understanding why the prescriptions we are providing, relating out-particles to in-particles, will preserve information.
	
Judicious attempts to map second-quantised Standard Model particles onto the radial first-quantised objects will produce novel insights that might, one day, enable us to find out what the primary source is of the complex structures of the Standard Model itself. In fact, the two transverse dimensions stay second-quantised\fn{The functions \(u_{\ell,m}\)  and \(p_{\ell,m}\) are the spherical harmonic transformations of the operators 
\(u(\tht,\vv)\) and \(p(\tht,\vv)\) and these are fields; this is why we talk of second quantisation in the transverse coordinates.}. Thus we see here the holographic principle at work.

As a first step, we arrive at a description of all states of a black hole in terms of states in a Hilbert space, where the spectrum of in-going particles is exactly as large as the spectrum of out-going particles, \emph{except for one very important point,} that we shall consider now. 

Our point is that, as was observed above, the operation to get the out-Hilbert space in  terms of the in-Hilbert space is the mathematical procedure of the Fourier transformation. The Fourier transformation does not distinguish the positive halves from the negative halves both in position space, \(u^\pm>0\) and momentum space, \(p^\mp>0\). Therefore, limiting ourselves only to region \(I\) of the black hole might not be correct. Can we resolve this problem?
\subsection{Keeping the even in- and the even out-wave functions separate}
Surprisingly, there is a beautiful solution. It is closely connected to our earlier consideration that the states on the 
Kruskal-Szekeres world frame
must be regarded as elements of the density matrix. More precisely, we address the entire region \(II\) as the space of bra states. The only states we shall accept on the entire KS coordinate frame is the set of `cloned twins':
\be |\psi_I,\,\psi_{II}\ket \ \ra\ |\psi\ket\bra\psi|\ , \ee
where \(|\psi\ket\) only consist of wave functions defined on region \(I\), and the states \(\bra\psi|\) are the same states, in bra notation, exclusively defined on region \(II\). This is not a trivial formalism at all, while many researchers are taking their refuge in much more entangled space-time metric frames. We claim that the approaches using more complex metric structures will perhaps be needed to describe further details at a later stage, but we first must understand and use the Ansatz that region \(II\) is a cloned version of region \(I\), as this is totally essential for recovering unitarity.
The situation differs from Rindler space\,\cite{Rindler.ref} in the sense that also the observer is  formally copied, see Figure \ref{clones.fig}. This restriction may be easier to accept by noting that the original Schwarzschild metric, \eqn{Schwmetric.eq}, already automatically exhibits every state only once, and \emph{there is only one observer in the Schwarzschild coordinate system.} See Fig.~\ref{clones.fig}.

The way this new picture affects our equations is that now we have a map of the in-going wave function  defined exclusively on the positive half of the entire future horizon (while the part where \(u^+<0\) is fixed as an exact copy of \(u^+>0\), and we demand in turn also the past horizon to be similar: the part where \(u^->0\) will be constructed as an exact copy of \(u^-<0\) (see also Fig.~\ref{SS-KS.fig}b). 

Eq.~\eqn{eq98.eq} relates the \emph{entire} future horizon to the entire past horizon. How does this change if the two halves of both horizons are restricted to being copies of the other halves? This happens to be easy. We write 
\be \psi_\inn(p)=\psi_\inn(-p)\ ,\quad \psi_\outt(u)=\psi_\outt(-u)\ ;\eel{eq910.eq}

This is not in conflict with the demand that the functions in \(u^-\) space are Fourier transforms of the ones in \(p^+\) space: the Fourier transform of an even function is an even function itself. Therefore, we may assume 
Eq.~\eqn{eq98.eq} to hold for positive \(u^\outt\) and \(p^\inn\) separately from the same equations for negative
 \(u^\outt\) and \(p^\inn\).

\subsection{Replacing the Fourier transform by an even function}
We are now in a position to formulate precisely the boundary condition that maps in-going energy-momentum that enters through the positive half of the future horizon, onto the states that return from half the past horizon, while all of this happens primarily in region \(I\); it is exactly copied in region \(II\).

The wave functions in \(p^-\) space are the Fourier transforms of the states in \(u^+\) space, even if we only take half of the available space, that is, on the Cauchy surfaces spanned by the \(u^+\) axis we keep only the wave functions that are \emph{even} in \(u^+\).  Then we find that also their Fourier transforms are even functions of \(p^-\). Proof:
\be %\psi_\outt^-(u)=\sqrt {\frac 1{2\pi}}
\psi^\outt(u)=\psi^\outt(-u)\ =\ \frac 1{\sqrt {2\pi}}\int_{-\infty}^\infty\dd p\,\ex{i\,p\,u}\,\psi^\inn(p)&=& 
\frac1{\sqrt{2\pi}}\int_0^\infty(\ex{ipu}+\ex{-ipu})\psi^\inn(p) =\nm\\ 
\sqrt {\frac 2{\pi}}\int_0^\infty\cos(pu)\,\psi^\inn(p)\ ,   \eel{eq103.eq}
which is its own inverse:
\be \psi^\inn(p)=\sqrt {\frac 2{\pi}}\int_0^\infty\cos(pu)\,\psi^\outt(u)\ ,   \eel{eq104.eq}

Exactly the same events happen in region \(II\).

This implies that the subsets of all even wave functions are transformed onto each other. Or: if the in going states form a set of quantum clones, then the out going states also form quantum clones automatically. In other words, even in these subsets the Fourier transform is a unitary transformation.
The Fourier transform is then effectively replaced by the transformation \eqn{eq104.eq}.

\newsecl{Discussion. Information and temperature}{infotemp.sec}
\begin{figure}\qqqquad\qquad
\widthfig{300pt}{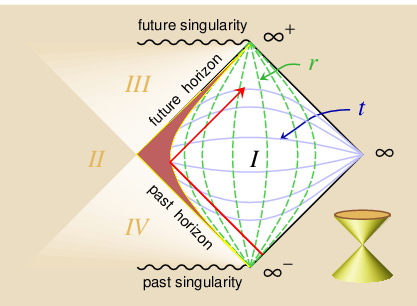}
\caption {\small The brick wall model bounces information carrying particles back out.\labell{brickwall.fig}}
\end{figure}

\subsection{Comparing the horizon with a brick wall or mirror}
All of the above, taken together, gives us an interesting glimpse of a theory where all information is passed on from in-going particles to out-going ones, \emph{via} unitary transmission operators, \eqn{eq103.eq}, \eqn{eq104.eq}. These unitary operators turn the horizons into effective mirrors.
 Note that, this turns our model into something very much resembling the `brick wall model', introduced long 
 ago\,\cite{GtH1985.ref} %21
 ; see Fig.~\ref{brickwall.fig}.

It is instructive to describe the sequences of operations that take proper care of the time dependence. At a given time \(\tau_0\), we describe all in-particles as wave functions \(|\psi^\inn_{\ell m}(u^+)\ket\), on the future event horizon \(u^+\), where \(u^+>0\). As time proceeds, \(u^+_\inn\) (and with it, \(p^{+\,\outt}\)), will decrease exponentially, so that the wave functions will collapse towards the origin. Their (modified) Fourier transforms, \(\psi_{\ell m}^\inn(p)\), see Eq.~\eqn{eq104.eq}, will expand exponentially in time, since the product \((pu)\) in this transformation then stays constant. Now Eq.~\eqn{eq98.eq} will relate \(p^\inn\) to \({u^\outt}\). Clearly, \(u^\outt\) will also expand. These are the out-particles, coming from the past horizon, doing their best, exponentially, to leave the scenery. Clearly, they contain all information that went in; nothing is lost -- apart from the non-geometrical information, as explained. 
\subsection{The transition to different time slices\labell{timeslices.sub}}
We apply this mechanism for a short time period. Now consider the next short time period. The particles that have been created before, will rapidly leave. They might still feel the forces from the previous period, but these are now too weak to have any effect. This implies that their spot in \((\ell, m)\) space is open for the next round.
New particles will arrive and the same action will be repeated.

Classically, all these wave functions will also continue into regions \(III\) and \(IV\), but we are not interested in that, since they will be invisible from the outside world.

There are also particles that may have entered from far away to merge with the future event horizon. As soon as they come too close to the horizon, they may become effectively invisible. In that case, we apply the inverse of our new transformation, to turn them into particles that leave. This is what we called the `firewall transformation' in Ref.~\cite{GtHfirewall.ref} %22
.

Note that this, finally, justifies what we postulated in the beginning, which is that we may omit particles that are either too close to the horizon or too far away. This set continually changes as we go from one time window to the next, keeping all our particles in the range where they are allowed (by obeying the constraint that their energies and momenta are negligible \emph{in this time frame}).
\subsection{Pure states and mixed states}

    Hawking\,\cite{Hawkdens.ref}, Giddings\,\cite{Giddings.ref}, and many others, conclude  rather prematurely that black holes must turn pure quantum states into mixed states. Here, we see nothing of the sort happening. \emph{We have been dealing with density matrices from the beginning.} Consider a pure quantum in-state, \(|\psi\ket=\psi(p^-_\inn)\). Its density matrix is \(\rr_\inn=|p^-_\inn\ket\bra p^-_\inn|\). Being a pure state implies that contributions of the form \(|\psi_1\ket\bra\psi_2|\), with \(\psi_2\ne\psi_1\), should not be there. This is exactly the statement that the states for positive \(p^-_\inn\) are the same as the ones for negative \(p^-_\inn\). Therefore: 

\emph{in-states of the form \(\psi(p^-)=\psi(-p^-)\) yield out-states of the same form: \(\psi(u^-)=\psi(-u^-)\), so that 
\(\psi(p^+)=\psi(-p^+)\).}

Note that, according to the transformation rules \eqn{eq103.eq} and \eqn{eq104.eq}, a change in the sign of \(p\) has the same effect as a change in the sign of \(u\). We must conclude that the complete Hilbert space to be worked with consists of the states \(\psi(u)\) with only positive values of \(u\), or equivalently, the states \(\psi(p)\) with only the positive values of \(p\). These transformations are unitary. Indeed, they map region \(I\) into region \(I\) and region \(II\) into region \(II\), as they should, from a physical point of view.
\subsection{Temperature}
This implies that the density matrices are spanned by the Hilbert space vectors \(|\psi\ket\,\bra\psi|\), so that, indeed, the states \(|\W\ket\) are density matrix elements. Their amplitudes for states with \(n\) particles are \(C^{st}\,e^{-n\pi\w}\), as they should be when the temperature is \(1/\pi\), not \(1/2\pi\), and consequently 
the Hawking temperature for a black hole with mass \(M\) must be
\be T=4\pi GM\eel{T2.eq} rather than Eq.~\eqn{Tqu.eq}.

There is an interesting argument explaining the relation between temperature and information. The behavior of a field theory at a given temperature \(1/\bet\), can be understood in terms of the functional integral for a statistical (thermal) system that extends for a stretch of time in the Euclidean 4-direction, with a length \(\bet\). We could continue the functional integral for another amount of \(\bet\) in Euclidean time, making the universe twice as big. This would make the total entropy of this field twice as big, since entropy simply counts the number of possible states (logarithmically). The total energy \(GM\), on the other hand, is fixed in relation with the black hole mass. Since
\be T\dd S=\dd M\ , \eel{entropy.eq}
the temperature \(T\) must increase by a factor 2 if entropy \(S\)  shrinks by a factor 2. 

Note that identifying region \(II\) with region \(I\) implies that, in the thermal functional integral, the variables in region \(II\) are fixed as soon as those in region \(I\) are, so that these cannot be included in the calculation of the total entropy. 
This argument can be elaborated more, but we merely mention it to show how the physical features of our theory hang together.

Strauss and Whiting\,\cite{StrWg.ref} %23
 report that they tried to comprehend what we are suggesting, and continued by testing the idea in the case of spherically symmetric dust clouds. But it should have been clear that the entire Hilbert space that we investigated, is dominated by non spherically symmetric states, and information retrieval is only conceivable if one considers all components of Hilbert space that can be involved (remember the heat bath in Subsection \ref{heatbath.sub}). It should have been obvious that black holes with spherically symmetric dust clouds added, will emit Hawking radiation that not only consists of such clouds.

Remarks such as those made by Strauss and Whiting will perhaps be useful if one is interested in rather drastic toy models, but it is relatively easy to consider the effects of all radial momenta \((\ell,m)\ne (0,0)\) combined.

\subsection{The conical singularity\labell{conical.sub}}
	Finally, there is the question whether the identification of regions \(I\) and \(II\) could generate a special type of singularity on the horizons. The answer is yes, at the points \(u^\pm\ra 0\), this identification can be removed by the transformation \((x,y)\leftrightarrow (x^2,y^2)\), since the operator \(\ZZ(2)\) mentioned earlier then becomes the identity operator, but a mild singularity at \(x=y=0\) then does result. In the equations \eqn{KS.eq}, we see that the time \(t\) goes to \(2t\), Elaborating this theme a bit further, we see that the temperature, which has dimension of an inverse time, is affected by a factor two when taking this conical singularity into account.

We thank N. Gaddam, U. G\"ursoy, N.~Groenenboom and S. Matur for useful discussions and remarks. Also N. Strauss and B. Whiting,  for writing up their considerations and findings\cite{StrWg.ref}, even if we do not always agree. It is the kind of discussions that may lead us into considering the next steps.

\end{document}